\documentclass[12pt]{article}
\usepackage{amssymb,eufrak,theorem,latexsym}

\newcommand{\nc}{\newcommand}
\nc{\la}{\lambda} \nc{\alf}{\alpha}
\nc{\tht}{\theta}  \nc{\be}{\beta}  \nc{\eps}{\epsilon}
\nc{\ga}{\gamma}  \nc{\Ga}{\Gamma}  \nc{\De}{\Delta}
\nc{\de}{\delta} \nc{\si}{\sigma}  \nc{\ka}{\kappa}
\nc{\om}{\omega}  \nc{\qq}{\quad\quad}
\nc{\nf}{\infty}   \nc{\dl}{\mathop{\smash{\cal L}}}
\nc{\ra}{\rightarrow}  \nc{\ol}{\overline}
\nc{\beq}{\begin{equation}}  \nc{\barr}{\begin{array}}
\nc{\earr}{\end{array}}
\nc{\eeq}{\end{equation}}
\nc{\beqa}{\begin{eqnarray}}  \nc{\dst}{\displaystyle}\nc{\pt}{\partial}
\nc{\eeqa}{\end{eqnarray}} \nc{\nnb}{\nonumber}
\nc{\bs}{\backslash}        \nc{\mbb}{\mathbb}
\nc{\brm}{\begin{remunerate}} \nc{\erm}{\end{remunerate}}
\nc{\nn}{\nonumber \\}
\nc{\p}[1]{{(\ref{#1})}}  \nc{\vareps}{\varepsilon}
\nc{\tb}{\tilde\beta_0}
\nc{\ts}{\tilde s}
\nc{\tth}{\tilde \theta}
\newcounter{muni}
\usecounter{muni}
\newtheorem{nth}{Proposition}

\newcommand{\stdec}{\mathop{\star}}  \nc{\lapdec}{\mathop{\Delta}}

\def\theequation{\arabic{section}.\arabic{equation}}
\newenvironment{remunerate}{\begin{list}{{\rm \arabic{muni}.}}
{\usecounter{muni}
\setlength{\leftmargin}{0pt}\setlength{\itemindent}{38pt}}}{\end{list}}
\begin{document}
\begin{titlepage}
\begin{flushright}
LPTHE 01-59 \\
hep-th/0110280 \\
October 2001
\end{flushright}
\vskip 1.0truecm
\begin{center}
{\large \bf $\mathbf{U(1) \times U(1)}$ QUATERNIONIC METRICS FROM HARMONIC SUPERSPACE}
\end{center} \vskip 1.0truecm
\centerline{\bf Pierre-Yves Casteill${}^{\;a,1}$, Evgeny
Ivanov${}^{\;b,a, 2}$, Galliano Valent${}^{\;a,3}$}
\vskip 1.0truecm
\centerline{${}^{a}$ \it
Laboratoire de Physique Th\'eorique et des Hautes Energies,}
\centerline{\it Unit\'e associ\'ee au CNRS URA 280,~Universit\'e Paris 7}
\centerline{\it 2 Place Jussieu, 75251 Paris Cedex 05, France}
\vskip5mm
\centerline{${}^{b}$\it Bogoliubov Laboratory of Theoretical
Physics, JINR,}
\centerline{\it Dubna, 141 980 Moscow region, Russia}

\vskip 1.0truecm  \nopagebreak

\begin{abstract}
\noindent We construct, using harmonic superspace and the quaternionic
quotient  approach, a quaternionic-K\"ahler extension of the most general two
centres hyper-K\"ahler  metric. It possesses $U(1)\times U(1)$ isometry,
contains as special cases the quaternionic-K\"ahler extensions of the Taub-NUT
and Eguchi-Hanson metrics and exhibits an extra one-parameter freedom which
disappears in the hyper-K\"ahler limit. Some emphasis is put on the relation
between this class of quaternionic-K\"ahler metrics and self-dual Weyl
solutions of the coupled Einstein-Maxwell equations. The relation between our
explicit results and the recent general ansatz  of Calderbank and Pedersen for
quaternionic-K\"ahler metrics with $U(1)\times U(1)$  isometries is traced in
detail.
\end{abstract}
\vfill {\it E-Mail:}\\ {\it 1) casteill@lpthe.jussieu.fr}\\ {\it 2)
eivanov@thsun1.jinr.ru, eivanov@lpthe.jussieu.fr}\\ {\it 3)
valent@lpthe.jussieu.fr}

\end{titlepage}

\section{Introduction}
Recently, there was a surge of interest in the explicit construction of
metrics for various classes of the hyper-K\"ahler (HK) and
quaternionic-K\"ahler (QK) manifolds, caused by the important role these
manifolds play in string theory (see, e.g., \cite{1}-\cite{gan}).
At present, there exist a few approaches to tackling this difficult
problem \cite{klr}-\cite{wrv2}. One of them proceeds from the
generic actions of bosonic nonlinear sigma models with the HK and QK target
manifolds \cite{gios1}-\cite{gios2}, \cite{gios3}-\cite{IvV}.

Such generic actions, respectively for the HK and QK sigma models, were
constructed in \cite{gios1,gio22,gios2} and \cite{gios3,bgio,gio,IvV} within
the harmonic superspace (HSS) method \cite{hss,CUP}, based on the renowned
one-to-one correspondence \cite{agf,bw} between the HK and QK manifolds on the
one hand, and global and local $N=2, d=4$ supersymmetries on the other. It was
proved in \cite{agf,bw} that the most general self-coupling of $N=2$ matter
supermultiplets (hypermultiplets) in the rigid or local $N=2$ supersymmetry,
necessarily implies, respectively, the HK or QK target geometry for  the
hypermultiplet physical bosonic fields. Conversely, any HK or QK bosonic sigma
model  can be lifted to a rigidly or locally $N=2$ supersymmetric nonlinear
sigma model. Most general off-shell actions for such $N=2$ sigma models were
constructed in \cite{gios2,gios3} in the framework of $N=2$ harmonic
superspace (HSS) \cite{hss} as the only one to offer such an opportunity. As
was proved in \cite{gios2,gio} starting from  the general definition
of HK or QK geometries as the properly constrained  Riemannian ones, the
corresponding analytic superfield Lagrangians of interaction have a nice
geometric interpretation as the HK or QK potentials. These are  the
fundamental objects of the HK and QK geometries (like the K\"ahler potential
in K\"ahler geometry). They encode the entire information about the local
properties of the relevant bosonic metric, in particular, about its
isometries. Then, based on the one-to-one correspondence  mentioned above, the
 generic HK and QK sigma model bosonic actions can be obtained simply by
discarding the  fermionic fields in the general harmonic superspace sigma
model actions. For the QK case such a generic bosonic action was constructed
in \cite{IvV}. The actions of physical bosons containing the explicit HK or QK
metric associated with the given harmonic potential appear  in general as the
result of elimination of infinite sets of auxiliary fields contained in the
off-shell hypermultiplet harmonic analytic superfields. This procedure amounts
to solving some differential equations on the internal sphere $S^2$
parametrized by the $SU(2)$ harmonic variables. It is a difficult problem in
general to solve such equations. However, as was shown in  \cite{giot,IvV}, in
the cases with isometries the computations can be radically
simplified  by using the harmonic superspace version of the HK \cite{klr,hklr}
 or QK \cite{gal1}-\cite{gal3} quotient constructions. One of the attractive
features of the HSS quotient is that it allows one, at all steps of
computation, to keep manifest the corresponding isometries of the metric which
come out as internal symmetries of the HSS sigma model Lagrangian with  a
transparent origin. It is especially interesting and tempting to apply this
method for the explicit calculation of new inhomogeneous QK metrics. Indeed,
whereas a lot of the HK metrics of this sort was explicitly constructed (both
in 4- and higher-dimensional cases, see, e.g., \cite{ksd}-\cite{Hi},
\cite{ggr}), not too many  analogous QK metrics are known to date.

In \cite{IvV}, using the HSS quotient techniques, we constructed QK extensions
of  the well-known \cite{egh} Taub-NUT and Eguchi-Hanson 4-dimensional HK
metrics and discussed  some their distinguished geometric features. In one or
another (though rather implicit)  form these  QK metrics already appeared in
the literature (see, e.g., \cite{gal2,iv1,To}) and  our detailed treatment
of them was a preparatory step to reveal  capacities of the  HSS approach for
working out more interesting and less known examples.

In \cite{gorv}, the double Taub-NUT HK metric  was derived from the HSS
approach  by directly solving the corresponding harmonic differential
equations. It turns out that the HSS quotient approach allows one to
reproduce the same answer much easier, and it nicely works as well in the QK
case, where solving similar harmonic equations would bear a much more
involved problem. In \cite{civ} we constructed a QK extension of the
double Taub-NUT metric using the HSS quotient approach.

The present paper is intended, on the one hand, to give the detailed proof of
some statements made in the letter \cite{civ} and to perform a further
comparison with the available ansatzes for QK metrics. On the other hand, we
demonstrate here that the HSS quotient approach suggests a further extension
of the class of explicit QK metrics presented in \cite{civ}. All of them
possess $U(1)\times U(1)$ isometry and are characterized by two additional free
parameters. In the HK limit they go over into a generalization of the standard
double Taub-NUT metric with two unequal ``masses'', one of the new parameters
being just the ratio of these  ``masses''. Another parameter does not show up
in the HK limit, but it proves essential at the non-vanishing contraction
parameter (Einstein constant). Thus we observe the existence of a
one-parameter class of non-equivalent QK metrics having the same HK limit.

In section 2 we remind the basic facts about the HSS action of generic QK
sigma model, as it was derived in \cite{IvV}. In section 3 we construct the
HSS quotient for the considered case of the QK double-Taub-NUT sigma model:
proceed from a sum of the HSS ``free'' actions of three $Q^+$
hypermultiplets (having the hyperbolic ${\mbb H}H^{3}$ manifold as the target
space) and then gauge  two common commuting one-parameter symmetries of these
actions by two non-propagating $N=2$ vector multiplets. The freedom in
embedding these two symmetries in the variety of symmetries of the ``free''
action  is characterized by two arbitrary constants which specify the most
general QK extension of the double Taub-NUT metric. \footnote{The QK metric
presented in \cite{civ} corresponds to the minimal case, when both extra
parameters are equal to zero.} The intermediate steps leading to the final
4-dimensional metrics are described in section 4. The metric is read off after
fixing the appropriate gauges and solving two sets of algebraic constraints
appearing as the equations of motion  for the auxiliary fields of the gauge
multiplets. In section  5 we bring the metrics into the final form.
Using the  Przanowski-Tod ansatz \cite{To,Pr}, we make an independent check
that the metrics are indeed self-dual Einstein. Several limiting cases are
also discussed. In section 6 we examine our metrics in the context of the
literature related to self-dual Einstein geometries \cite{Pe}-\cite{cp},
including  Flaherty's equivalence to the (self-dual Weyl) solutions of the
coupled Einstein-Maxwell equations \cite{Fl}.

Just after publication of our letter \cite{civ} reporting the
construction of a QK extension of the double Taub-NUT metric in the HSS
approach, Calderbank and Pedersen \cite{cp}  have obtained the exact
linearization of any four-dimensional QK metric with two commuting Killing
vectors.
After a short review of their results in section 6.5, we give the precise
relation between their coordinates and ours.

\setcounter{equation}{0}
\section{The generic HSS action of QK sigma models}
In \cite{IvV} the generic action of QK sigma models with $4n$ dimensional
target manifold of physical bosons was obtained as a pure bosonic part of the
general off-shell HSS action of $n$ self-interacting matter hypermultiplets
coupled to the so-called principal version of $N=2$ Einstein supergravity
\cite{gios3}. The gauge multiplet of the latter, in the language of $N=2$
conformal SG, consists of the $N=2$ Weyl multiplet (24 + 24 off-shell
components), the compensating vector multiplet (8+8 off-shell components)  and
the compensating hypermultiplet ($\infty + \infty$ off-shell components). It
is the only version which admits the most general hypermultiplet matter
self-couplings and thus, in accord with the theorem of \cite{bw}, the most
general QK metric in the sector of physical bosons. The matter and
compensating hypermultiplets are described by the   superfields
$Q^{+}_r(\zeta)$ and $q^{+}_a(\zeta)$, $r= 1,\ldots , 2n$, $a=1,2$, given on
the harmonic analytic $N=2$ superspace  \beq (\,\zeta \,) = (\,x^m,
\theta^{+\mu}, \bar\theta^{+\dot\mu}, u^{+ i}, u^{- k}\,)~,  \label{anal}
\eeq where the coordinates $u^{+ i}, u^{- k}, \;u^{+i}u^-_i = 1,\; i,k = 1,2$,
are the $SU(2)/U(1)$ harmonic variables. These superfields obey the
pseudo-reality conditions    \beq (a)\; Q^{+ r} \equiv \widetilde{(Q^+_r)} =
\Omega^{rs}Q^{+}_s~, \quad  (b) \;q^{+ a} \equiv \widetilde{(q^+_a)} =
\epsilon^{ab}q^{+}_b~, \label{real}
\eeq
where $\Omega^{rs}$ and
$\epsilon^{ab}$ ($\epsilon^{12} = -\epsilon_{12} = -1$) are the skew-symmetric
constant $Sp(n)$ and $Sp(1) \sim SU(2)$ tensors. The generalized  conjugation
$\;\widetilde{}\;$ is the product of the ordinary complex conjugation and a
Weyl reflection of the sphere $S^2 \sim SU(2)/U(1)$ parametrized by $u^{\pm
i}$. The superspace \p{anal} is real with respect to this generalized
conjugation which acts in the following way on the superspace coordinates :
$$
\widetilde{x^m} = x^m~, \quad \widetilde{\theta^{+\mu}} =
\bar\theta^{+\dot\mu}~, \quad  \widetilde{\bar\theta^{+\dot\mu}} = -
\theta^{+\mu}~, \quad \widetilde{u^{\pm}_i} = u^{\pm i}~, \quad
\widetilde{u^{\pm\,i}} = -u^{\pm}_{ i}~.
$$
In the QH sigma model action to be given below we shall need to know only the
bosonic components in the $\theta$-expansion of the above superfields:
\beqa q^{+a}(\zeta) &=& f^{+ a}(x,u) + i(\theta^+\sigma^m\bar\theta^+)A^{-
a}_{m}(x,u) + (\theta^+)^2(\bar\theta^+)^2 g^{-3 a}(x,u) \nn  Q^{+ r}(\zeta)
&=& F^{+ r}(x,u) + i(\theta^+\sigma^m\bar\theta^+)B^{- r}_{m}(x,u) +
(\theta^+)^2(\bar\theta^+)^2 G^{-3 r}(x,u) \label{thetaexp1}
\eeqa
(possible terms $\sim (\theta^+)^2$ or $\sim (\bar\theta^+)^2$ can be shown to
fully drop out from the final action and so can be discarded from the very
beginning). The component fields still have general harmonic expansions off
shell. The physical bosonic components $F^{ri}(x), f^{ai}(x)$ are defined as
the lowest components in the harmonic expansions of $F^{+ r}(x, u), f^{+ a}(x,
u)$
\beqa && F^{+r}(x,u) = F^{ri}(x)u^+_i + \cdots~, \quad  f^{+a}(x,u) =
f^{ai}(x)u^+_i + \cdots~, \nn
&&\overline{(F^{ri}(x))} =
\Omega_{rs}\epsilon_{ik}F^{sk}(x)~, \; \overline{(f^{ai}(x))} =
\epsilon_{ab}\epsilon_{ik}f^{bk}(x)~.
\eeqa
Further details can be found in \cite{IvV} and \cite{hss}.

The bosonic QK sigma model action derived in \cite{IvV} consists of the two
parts
\beqa
S_{QK} &=& {1\over 2} \int d\zeta^{(-4)}\left\{-q^+_a {\cal D}^{++}q^{+a} +
\frac{\kappa^2}{\gamma^2} (u^-_i q^{+i})^2 \left[ Q^+_r {\cal D}^{++}Q^{+r} +
L^{+4}(Q^+, v^+, u^-) \right]\right\} \nn
&& - {1\over 2\kappa^2}\int d^4x \left[D(x) +
{\cal V}^{m\,ij}(x){\cal V}_{m\,ij}(x)\right] \equiv S_{q,Q}+ S_{SG}~.
\label{QKact}
\eeqa
Here, $d \zeta^{(-4)} = d^4x d^2\theta^+ d^2\bar\theta^+ du$ is the
measure of integration over \p{anal}, the covariant harmonic derivative
${\cal D}^{++}$ is defined by
\beq
{\cal D}^{++} = D^{++} +
(\theta^+)^2(\bar\theta^+)^2\,
\{\; D(x)\,\partial^{--} + 6\,{\cal V}^{m\,(ij)}(x)u^-_iu^-_j \,\partial_m \;
\}~, \label{substit}
\eeq
with $D^{++} = \partial^{++} - 2i\theta^+\sigma^m\bar\theta^+\,\partial_m$,
$\partial^{\pm\pm} = u^{\pm i}/\partial u^{\mp i}$, the non-propagating
fields $D, {\cal V}^{ij}_m = {\cal V}^{ji}_m$ are inherited from the $N=2$
Weyl multiplet, $\kappa^2 \;([\kappa] = -1)$ is the Einstein constant (or,
from the geometric standpoint, the parameter of contraction to  the HK case),
$\gamma \; ([\gamma] =-1)$ is the sigma model constant (chosen equal to 1
from now on), and the ``target'' harmonic variable $v^{+ a}$ is defined by
\beq  v^{+ a} = \frac{q^{+a}}{u^-_iq^{+i}} = u^{+ a} - \frac{u^+_iq^{+i}}{u^-_i
q^{+i}} \,u^{- a}~, \quad v^{+a}u_a^- = 1~. \label{defv}
\eeq
The function $L^{+4}(Q^+, v^+, u^-)$ is the analytic QK potential, the object
which encodes the full information about the relevant QK metric.

The action \p{QKact} possesses a local $SU(2)$ invariance, the remnant
of the $N=2$ supergravity gauge group, with ${\cal V}^{ij}_m(x)$ as the gauge
field. The precise form of the $SU(2)_{loc}$ transformations leaving
the $S_{q,Q}$ part of \p{QKact} invariant can be inferred from the
realization of the group  of $N=2$ conformal SG as restricted diffeomorphisms
of the analytic superspace \p{anal} \cite{confSG}. This can be achieved  by
fixing a WZ gauge for the Weyl multiplet and neglecting all its field
components besides $D(x)$, ${\cal V}^{ik}_m(x)$, $e^a_m(x) \rightarrow
\delta^a_m$ and all the residual gauge invariance parameters besides the
$SU(2)_{loc}$ one $\lambda^{ik}(x)=\lambda^{ki}(x)$. These transformations
read \footnote{They were not explicitly given in \cite{IvV} and earlier
papers on the subject.}
\beqa \delta u^+_i &=& \Lambda^{++} u^{-}_i~,
\qquad \delta u^-_i = 0~, \nn \Lambda^{++} &=& \lambda^{++} + 2i
\theta^+\sigma^m\bar\theta^+\partial_m\lambda^{+-} -
(\theta^+)^2(\bar\theta^+)^2\left(\Box \lambda^{--} +4
{\cal V}^{--m}\partial_m\lambda^{--} \right. \nn
&& \left. - \,  2{\cal V}^{+-m}\partial_m \lambda^{--} -
\lambda^{--} D\right)~, \nn
\delta \theta^{+\mu} &=& \lambda^{+-}\theta^{+
\mu} - i (\theta^+)^2 (\sigma^m\bar\theta^{+})^\mu\partial_m\lambda^{--}
\equiv \lambda^{+\mu}(\zeta)~,
\; \delta \bar\theta^{+\dot\mu} = \widetilde{(\delta\theta^{+\mu})} =
\bar\lambda^{+\dot\mu}(\zeta)~, \nn
\delta x^{m} &=& -2i
\theta^{+}\sigma^m\bar\theta^{+} \lambda^{--} + 6
(\theta^+)^2(\bar\theta^+)^2 {\cal V}^{--\,m}\lambda^{--} \equiv
\lambda^m(\zeta)~, \label{coordSU2} \\
\delta {\cal D}^{++} &=& -\Lambda^{++}D^0~, \quad D^0 =
u^{+ i}\frac{\partial}{\partial u^{+ i}} - u^{- i}\frac{\partial}{\partial
u^{-i}} +  \theta^{+ \mu}\frac{\partial}{\partial \theta^{+ \mu}}
+\bar\theta^{+ \dot\mu}\frac{\partial}{\partial \bar\theta^{+\dot\mu}}~,
\label{DSU2} \\
\delta q^{+ a}(\zeta) &\simeq & q^{+ a}{}'(\zeta') - q^{+a}(\zeta) = -{1\over
2}\Lambda(\zeta) q^{+a}(\zeta)~, \label{qSU2} \\
\delta Q^{+ r}(\zeta)
&\simeq & Q^{+ r}{}'(\zeta') - Q^{+r}(\zeta) = 0~, \label{QSU2} \\
\Lambda(\zeta) &=& \partial_m\lambda^m + \partial^{--}\Lambda^{++} -
\partial_{+\mu}\lambda^{+\mu} - \partial_{+\dot\mu}\bar\lambda^{+\dot\mu}~.
\label{Lambda}
\eeqa
Here
$$
\lambda^{\pm\pm} = \lambda^{ik}(x)u^{\pm}_iu^{\pm}_k~, \;
\lambda^{+-} = \lambda^{ik}(x)u^{+}_iu^{-}_k~, \; {\cal V}^{--}_m =
{\cal V}^{ik}_m(x)u^-_iu^-_k~, \;  {\cal V}^{+-}_m =
{\cal V}^{ik}_m(x)u^+_iu^-_k~.
$$
To these transformations one should add the transformation laws
of the fields $D(x)$ and ${\cal V}^{ik}_m(x)$
\beq
\delta^* D(x) = 2\partial_m\lambda^{ik}(x){\cal V}^m_{ik}(x)~, \quad
\delta^* {\cal V}^{ik}_m(x) = -\partial_m\lambda^{ik}(x) + 2
\lambda^{(i}_j(x){\cal V}_m^{k)j}(x)~, \label{VDSU2}
\eeq
which uniquely follow from the transformation law \p{DSU2}.\footnote{Though
looking rather involved, the transformations \p{coordSU2}-\p{VDSU2} can be
straightforwardly checked to be closed, with the Lie bracket parameter
$\lambda^{ik}_{br} = \lambda_2^{il}\lambda_{1\,l}^{\;\;k} -
\lambda_1^{il}\lambda_{2\,l}^{\;\;k}$.}  It is easy to see that the $S_{SG}$
part of \p{QKact} is invariant  under \p{VDSU2}, implying the $SU(2)_{loc}$
invariance of the full action \p{QKact}. Note that the QK potential
$L^{+4}(Q^+, v^+, u^-)$ in \p{QKact} is $SU(2)_{loc}$ invariant because its
arguments $Q^{+r}, v^{+a}$ and  $u^{- i}$ behave as scalars under the above
transformations. The transformations \p{coordSU2}-\p{Lambda} entail the
following simple $SU(2)_{loc}$ transformation rules for the lowest components
$f^{+a}(x,u)$, $F^{+r}(x,u)$ in the $\theta$-expansion \p{thetaexp1}  \beq
\delta^* f^{+a} = \lambda^{+-}f^{+a} - \lambda^{++}\partial^{--}f^{+a}~, \quad
 \delta^* F^{+r} = - \lambda^{++}\partial^{--}F^{+r}~. \label{fFSU2} \eeq

The procedure of obtaining the QK metric from the action \p{QKact}
goes through a few steps. First one integrates over $\theta$s in
$S_{q,Q}$, then varies with  respect to the non-propagating  fields
$g^{-3 a}(x,u)$, $G^{-3 r}(x,u)$,  $A^{- a}_{m}(x,u)$,
$B^{- r}_{m}(x,u)$, $D(x)$ and ${\cal V}^{ij}_m(x)$, solve the resulting
non-dynamical equations and substitute the solution back into  \p{QKact}, thus
expressing everything in terms of the physical components $f^{ai}(x)$ and $F^{r
i}(x)$. Varying with respect to $D(x)$ and ${\cal V}^{ik}_m(x)$ yields the
important constraint relating $f^{+a}$ and $F^{+r}$:
\beq \label{constr}
\int du \left[ f^{+a}\partial^{--}f^+_a -\kappa^2(u^-f^+)^2\,
F^{+r}\partial^{--}F^{+}_r \right] ={1\over \kappa^2}
\eeq
and the general expression for ${\cal V}^{ik}_m$ in terms of
the hypermultiplet fields
\beq
{\cal V}^{ik}_m(x) = 3\kappa^2 \int du\,u^{-i}u^{-k} \left[ f^{+a}\partial_m
f^+_a -\kappa^2(u^-f^+)^2\, F^{+r}\partial_mF^{+}_r \right]~. \label{exprV}
\eeq
As the next step, one fixes a gauge with
respect to the $SU(2)_{loc}$ transformations defined above. Most
convenient is the gauge leaving only the singlet part in $f^{ai}(x)$
\beq f^{i}_a(x) =
\delta^{i}_a\, \omega(x) \label{gauge}
\eeq
(in what follows, we shall permanently use just this gauge). Finally, using
the constraint \p{constr}, one expresses $\omega$ in terms of
$F^{ri}(x)$, substitutes this expression into the action and reads  off the
QK metric on the $4n$ dimensional target space parametrized by  $F^{ri}(x)$.

An essential assumption is that $\omega$ is a constant in the flat
(hyper-K\"ahler) limit which is achieved by putting altogether
\beq
\vert \kappa \vert \omega = 1~,
\eeq
and then setting
\beq
\kappa = 0.
\eeq
Note that in order to approach the HK limit in \p{QKact} in the unambiguous
way, one should firstly eliminate the non-propagating field ${\cal
V}^{ik}_m(x)$ by its algebraic equation of motion and also perform varying
with respect to the auxiliary field $D(x)$. Taking into account that the
composite field ${\cal V}^{ik}_m(x) \sim O(\kappa^2)$ and $q^{+a}
\rightarrow u^{+a}\vert \kappa \vert^{-1}$ in the HK limit, one observes
that any  dependence on $q^{+a}, D$ and ${\cal V}^{ik}_m$ disappears in this
limit, and \p{QKact} goes into the HSS action of generic HK sigma model of $n$
hypermultiplets $Q^{+r}$ ($r = 1,...,2n$) \cite{gio22,gios2}. The constraint
\p{constr} becomes just the identity $1=1$. Another possibility is to
remove the fields $D(x), {\cal V}^{ik}_m(x)$ from \p{QKact} by equating
them to zero. In this case one reproduces the HSS action of the most general
conformally-invariant  HK sigma model with $n+1$ hypermultiplets
\cite{CUP,IvV,Ket} (the former compensator $q^{+a}(\zeta)$ enters it on equal
footing with other hypermultiplets). One can reverse the argument, i.e. start
from such HK sigma model action and reproduce the QK sigma model one
\p{QKact} by coupling the HK action to the non-propagating fields $D(x)$
and ${\cal V}^{ik}_m(x)$ in order to restore the local $SU(2)$ symmetry and
to be able to remove the remaining (non-gauge) bosonic degree of freedom in
$f^{+a}$ by  the constraint \p{constr}. This is the content of the so-called
``$N=2$ superconformal quotient'' approach to the construction of
$4n$-dimensional QK manifolds from the $4(n+1)$-dimensional HK ones
\cite{nider,swa,gal4,wrv1,wrv2}. In what follows we  shall not need to resort
to such an interpretation and shall proceed  from the general QK sigma model
action \p{QKact}.

\setcounter{equation}{0}
\section{QK extensions of the ``double Taub-NUT'' sigma model from HSS
quotient}

As already mentioned, on the road to the explicit QK metrics
one needs to solve the differential equations on $S^2$ for $f^{+a}(x,u)$,
$F^{+r}(x,u)$ which follow by varying the QK sigma model action with respect
to the non-propagating fields $g^{-3 a}(x,u)$ and $G^{-3r}(x,u)$. No regular
methods of solving such nonlinear equation  are known so far, and this can
(and does) bear some troubles in general. However, in a number of
interesting examples there is a way around this difficulty, the HSS quotient
method (it should not be confused with the ``superconformal quotient''
mentioned in the end of the previous section). It can be applied  both in the
HK \cite{giot} and QK \cite{bgio,IvV} cases. In it, one proceeds from  a
system of several ``free'' hypermultiplets (with $L^{+4} = 0$ in  \p{QKact},
which corresponds to a ${\mbb H}H^{n} \sim Sp(1,n)/Sp(1)\times Sp(n)$ sigma
model) and gauges some symmetries of this system in the analytic superspace by
non-propagating $N=2$ vector multiplets represented by the gauge superfields
$V^{++}(\zeta)$ (once again, only  bosonic components of these superfields
are of relevance). In one of possible  gauges these superfields can be fully
integrated out, producing a  non-trivial QK (or HK) potential $L^{+4}$ with
the necessity to solve nonlinear harmonic equations. But in another gauge
(Wess-Zumino gauge) the harmonic equations remarkably become  {\it linear}
and can be easily solved. All the nonlinearity in this gauge proves to be
concentrated  in nonlinear  algebraic constraints on the hypermultiplet
physical fields.  These constraints are enforced by the auxiliary fields of
vector multiplets as Lagrange multipliers. They are much easier to solve as
compared to the differential equations on $S^2$. This allows one to get the
explicit form of the QK (or HK) metric at cost of a comparatively  little
effort.

In \cite{IvV}, we exemplified the HSS quotient approach by QK extensions of
the Taub-NUT and Eguchi-Hanson (EH) metrics. Here we elaborate on a more
interesting and non-trivial case of the QK nonlinear sigma
model generalizing the HK model with the ``double Taub-NUT'' target
manifold.  The HSS action of the latter model was proposed in \cite{giot},
and  the relevant HK metric was directly computed in \cite{gorv} (it belongs to
the class of  two-center ALF metrics, with the triholomorphic $U(1)\times
U(1)$ isometry). Here we construct, using the HSS QK quotient method, the QK
sigma model action going into that of \cite{giot,gorv} in the HK limit.
We find an interesting degeneracy suggested by the QK quotient: there
is a one-parameter family of the QK metrics, all having $U(1)\times U(1)$
isometry and reproducing the double Taub NUT metric in the HK limit. More
general QK action contains one more parameter which survives in the HK limit
and corresponds to a generalization of the double Taub NUT metric by
non-equal ``masses'' in its two-centre potential.

\subsection{Minimal QK double-Taub-NUT HSS action}
The actions we wish to construct have as their ``parent action'' the QK
action  including three hypermultiplet superfields of the type $Q^{+r}$ with
the vanishing $L^{+4}$.  So it corresponds to the ``flat'' QK manifold ${\mbb
H}H^{3} \sim Sp(1,3)/Sp(1)\times Sp(3)$. For our specific purposes we relabel
this superfield triade as
\beq
Q^{+ a}_A, \;\;g^{+ r}~, \qquad a=1,2; \; r = 1,2; \;A= 1,2.
\eeq
The indices $a$ and $r$ are the doublet indices of two
(initially independent) Pauli-G\"ursey type $SU(2)$ groups realized on
$Q^{+}$ and $g^{+}$, the index $A$ is an extra $SO(2)$ index. Each of these
three superfields satisfies the pseudo-reality condition (\ref{real}).

We wish to end up with a 4-dimensional quaternionic metric. So,
following the general strategy of the quotient method, we need to gauge {\it
two commuting} one-parameter ($U(1)$) symmetries of this action. In this case
the total number of algebraic constraints and residual gauge invariances in
the WZ gauge is expected to be just 8, which is needed for reducing the
original 12-dimensional physical bosons target space to the 4-dimensional one.
These $U(1)$ symmetries should be commuting, otherwise their gauging would
entail gauging the symmetries appearing in their commutator. This would
result in further constraints trivializing the theory.

The selection of two commuting symmetries to be gauged and the form of
the final gauge-invariant HSS action are to a great extent specified by the
natural requirement that the resulting action has two different limits
corresponding to the earlier considered HSS quotient actions of  the QK
extensions of Taub-NUT and Eguchi-Hanson metrics \cite{IvV}. The simplest
gauged action $S_{dTN}$ which meets this demand is
\beq S_{dTN} = {1\over 2} \int d\zeta^{(-4)}{\cal
L}^{+4}_{dTN}  - {1\over 2\kappa^2}\int d^4x \left[D(x) +
{\cal V}^{m\,ij}(x){\cal V}_{m\,ij}(x)\right]~, \label{dTN1}
\eeq
where
\beqa   {\cal
L}^{+4}_{dTN} &=& -q^+_a{\cal D}^{++}q^{+a} +  \kappa^2 (u^-\cdot q^+)^2\left[
 Q^{+}_{rA}{\cal D}^{++}Q^{+r}_{A} + g^+_r{\cal D}^{++}g^{+r} \right.\nn  &&
\left.  + \, W^{++}\left( Q^{+a}_AQ^+_{a B}\epsilon_{AB} - \kappa^2
c^{(ij)}g^+_ig^+_j + c^{(ij)}v^{+}_iv^+_j \right) \right. \nn && \left.+
\,V^{++}\left( 2(v^+\cdot g^+) - a^{(rf)}Q^+_{rA}Q^+_{fA}\right) \right]
\label{action2}
\eeqa
and the second term ($S_{SG}$) is common for all QK sigma model actions. In
\p{action2},  $V^{++}(\zeta)$ and $W^{++}(\zeta)$ are two analytic gauge
abelian  superfields, $c^{(ij)}$ and $a^{(rm)}$ are two sets of
independent $SU(2)$ breaking parameters satisfying the pseudo-reality
conditions
\beq \overline{(c^{(ij)})} = \epsilon_{ik}\epsilon_{jl}c^{(kl)}~,
\quad \overline{(a^{(rm)})} = \epsilon_{rn}\epsilon_{ms}a^{(ms)}~.
\eeq
The Lagrangian can be checked to be invariant
under the following two commuting gauge $U(1)$ transformations, with the
parameters $\varepsilon(\zeta)$ and $\varphi(\zeta)$:\footnote{To avoid a
possible confusion, let us recall that the original general QK sigma model
action \p{QKact} contains a dimensionful sigma model constant
$\gamma$, $[\gamma] = -1$, which  we have put equal to 1 for convenience.
Actually, it is present in an implicit form in the appropriate places of eq.
\p{action2} and subsequent formulae, thus removing an apparent discrepancy in
the dimensions of various  involved quantities. From now on, we assign the
following dimensions to the basic involved objects and the gauge
transformation  parameters (in mass units): $[q] = [Q] = 1$, $[W^{++}] = 0$,
$[V^{++}] = 1$, $[c] = 2$, $[a] = -1$, $[\varepsilon ] = 0$, $[\varphi] = 1$.
With this choice, $\gamma $ nowhere re-appears on its own right.}
\beqa
&&\delta_{\varepsilon} Q^{+r}_A = \varepsilon \left[ \epsilon_{AB} Q^{+r}_B -
\kappa^2 c^{ij}v^+_iu^-_j Q^{+r}_A \right]~, \;\;\delta_\varepsilon g^{+r} =
\varepsilon \kappa^2 \left[ c^{(rn)} g^+_n -  c^{ij}v^+_iu^-_j g^{+ r}
\right]~, \nn  && \delta_\varepsilon q^{+ a} = \varepsilon \kappa^2
c^{(ab)}q^+_b~, \;\; \delta_\varepsilon W^{++} = {\cal D}^{++}\varepsilon~,
\label{eps} \eeqa
\beqa
&&\delta_\varphi Q^{+r}_A = \varphi a^{(rb)}Q^{+}_{b\,A}- \varphi \kappa^2
(u^-\cdot g^+) Q^{+r}_A~, \;\; \delta_\varphi g^{+r} = \varphi \left[ v^{+r}
- \kappa^2  (u^-\cdot g^+) g^{+r} \right]~,  \nn
&& \delta_\varphi q^{+ a} = \varphi \kappa^2 (u^-\cdot q^+)g^{+ a}~, \;\;
\delta_\varphi V^{++} = {\cal D}^{++}\varphi~. \label{phi}
\eeqa
This gauge freedom  will be fully fixed at the end. The only surviving
global symmetries of the action will be two commuting $U(1)$. One of them comes
from  the Pauli-G\"ursey $SU(2)$ acting on $Q^{+a}_A$ and broken by the
constant triplet $a^{(bc)}$. Another $U(1)$ is the result of breaking of the
$SU(2)$ which uniformly rotates the doublet indices of harmonics and those of
$q^{+a}$ and $g^{+r}$. It does not commute with supersymmetry (in the full
$N=2$ supersymmetric version of \p{action2}) and forms the diagonal subgroup
in the product of three independent  $SU(2)\,$s realized on these quantities
in the ``free'' case; this product gets broken  down to the diagonal $SU(2)$,
and further to $U(1)$, due to the presence of explicit harmonics and constants
$c^{(ik)}$ in the interaction terms in \p{action2}. These two $U(1)$
symmetries are going to be isometries of the final QK metric, the first one
becoming triholomorphic in the HK limit. The fields $D(x)$ and ${\cal
V}^{(ik)}_m(x)$ are inert under any isometry (modulo some rotations in the
indices $i, j$ after fixing the gauge \p{gauge}), and so are ${\cal D}^{++}$
and the $S_{SG}$ part of \p{dTN1}. The harmonics $v^{+a}$, as follows from
their definition \p{defv}, undergo some appropriate transformations induced
by those of $q^{+a}$ in \p{eps} and \p{phi}. Note that the presence of
the $g$-field term in the supercurrent (Killing potential) to which $W^{++}$
couples in \p{action2}, in parallel with the $v^+_i$ term (becoming the
Fayet-Iliopoulos term in the HK limit), is required for ensuring the invariance
of this supercurrent under the $\varphi$ gauge transformations. This in turn
implies the non-trivial transformation property of $g^{+r}$ under the
$\varepsilon $ gauge group in \p{eps}. In the HK limit the $g$-field term
drops out and $g^{+r}$ becomes inert under the $\varepsilon $ transformations.

By fixing the appropriate broken $SU(2)$ symmetries in \p{action2}, we
can leave only one real component in each of the $SU(2)$ breaking vectors
$a^{rf}$ and $c^{ik}$. Thus the relevant QK metric is characterized by three
real parameters: two $SU(2)$ breaking ones and the Einstein constant
$\kappa^2$. The $SU(2)$ breaking parameters survive in the HK limit.
\vspace{0.4cm}

\noindent\underline{\it The QK EH and Taub-NUT sigma model limits}\\

It is easy to see that the action \p{dTN1}, \p{action2} is indeed a
generalization of the HSS quotient actions describing QK extensions
of the  EH and Taub-NUT sigma models.

Putting $g^{+r} = a^{(rm)}= 0$ yields
the QK EH action as it was given in \cite{bgio,IvV}:
\beqa
{\cal L}^{+4}_{dTN}\; \Rightarrow  \; {\cal L}^{+4}_{EH} &=& -q^+_a{\cal
D}^{++}q^{+a}  +  \kappa^2 (u^-\cdot q^+)^2\left[Q^{+}_{rA}{\cal
D}^{++}Q^{+r}_{A} \right.\nn
&&\left.  + \, W^{++}\left( Q^{+a}_AQ^+_{a
B}\epsilon_{AB} + c^{(ij)}v^{+}_iv^+_j \right) \right]~. \label{EH}
\eeqa

Putting $Q^{+ a}_2 =  c^{(ik)} = 0$ yields the QK Taub-NUT action
\cite{IvV} \beqa
{\cal L}^{+4}_{dTN}\; \Rightarrow  \; {\cal L}^{+4}_{TN} &=& -q^+_a{\cal
D}^{++}q^{+a}  +  \kappa^2 (u^-\cdot q^+)^2\left[g^{+}_{r}{\cal
D}^{++}g^{+r}+Q^+_{1r}{\cal D}^{++}Q_1^{+r} \right.\nn
&&\left.  + \,V^{++}\left( 2(v^+\cdot g^+) - a^{(rf)}Q^+_{r1}Q^+_{f1}\right)
\right]~. \label{TN}
\eeqa
\vspace{0.4cm}

\noindent\underline{\it The HSS action with $g^{+r}$ eliminated}\\

Representing $g^{+r}$ as
$$
g^{+r} = (u^-\cdot g^+)v^{+r} - (v^+\cdot g^+)u^{-r}~,
$$
fixing  the gauge with respect to the $\varphi$
transformations by the  condition
$$(u^-\cdot g^+) = 0,$$
varying with respect to the non-propagating  superfield $V^{++}$ and
eliminating altogether $(v^+\cdot g^+)$ by the resulting algebraic
equation,
$$ (v^+\cdot g^+) \equiv L^{++} = {1\over
2}a^{rf}Q^+_{rA}Q^+_{fA}~, $$
we arrive at the following equivalent form of \p{action2}, with only two
matter hypermultiplets $Q^{+a}_A$ being involved
\beqa   {\cal
L}^{+4}_{dTN} &=& -q^+_a{\cal D}^{++}q^{+a} +  \kappa^2 (u^-\cdot q^+)^2\left[
 Q^{+}_{rA}{\cal D}^{++}Q^{+r}_{A} + L^{++}L^{++} \right.\nn  &&
\left.  + \, W^{++}\left( Q^{+a}_AQ^+_{a B}\epsilon_{AB} - \kappa^2
c^{(ij)}u^-_iu^-_j L^{++}L^{++} + c^{(ij)}v^{+}_iv^+_j \right) \right]~.
\label{action20}
\eeqa
In  the HK limit $\kappa^2 \rightarrow 0$ ($q^{+a} \rightarrow
\vert\kappa\vert^{-1}u^{+a}$, $\vert\kappa\vert (u^-\cdot q^+) \rightarrow 1$)
the corresponding action goes into the HSS action  describing the double
Taub-NUT  manifold \cite{giot,gorv}.\footnote{For the precise correspondence
one should choose $a^{12} = ia, a^{11} = a^{22} =0$ by appropriately fixing the
frame with respect to the broken Pauli-G\"ursey $SU(2)$ symmetry of $Q^{+r}$.}
Thus \p{dTN1}, \p{action2}  is the natural QK generalization of the action of
\cite{giot,gorv} and therefore  the relevant  metric is expected to be a QK
generalization of the double Taub-NUT HK  metric. We shall calculate it and
its some generalizations in the next sections by choosing another, Wess-Zumino
gauge in the relevant gauged QK sigma model actions.

\subsection{Generalizations}
In order to better understand the symmetry structure of the action \p{action2}
and to construct its generalizations, let us make the field redefinition
\beq
\hat{Q}^{+a}_A = \vert \kappa \vert (u^-\cdot q^+) Q^{+ a}_A~, \quad
\hat{g}^{+r} = \vert \kappa \vert (u^-\cdot q^+) g^{+ r}_A~. \label{hat}
\eeq
In terms of the redefined superfields, eqs. \p{action2}, \p{eps} and \p{phi}
are simplified to
\beqa
{\cal L}^{+4}_{dTN} &=& -q^+_a{\cal
D}^{++}q^{+a} +  \hat{Q}^{+}_{rA}{\cal
D}^{++}\hat{Q}^{+r}_{A} + \hat{g}^+_r{\cal D}^{++}\hat{g}^{+r} \nn
&& + \, W^{++}\left[ \hat{Q}^{+a}_A\hat{Q}^+_{a B}\epsilon_{AB} -
\kappa^2 c^{(ij)}\left(\hat{g}^+_i\hat{g}^+_j - q^{+}_iq^+_j \right)\right]
\nn
&& + \,V^{++}\left[ 2\vert \kappa \vert \left(q^+\cdot \hat{g}^+\right) -
a^{(rf)}\hat{Q}^+_{rA}\hat{Q}^+_{fA}\right]~, \label{action22}
\eeqa
\beqa
&&\delta_\varepsilon \hat{Q}^{+r}_A = \varepsilon \,\epsilon_{AB}
\hat{Q}^{+r}_B~, \;\;\delta_\varepsilon \hat{g}^{+r} = \varepsilon\, \kappa^2
c^{(rn)} \hat{g}^+_n~, \;\;\delta_\varepsilon q^{+ a} = \varepsilon \kappa^2
c^{(ab)}q^+_b~,   \label{eps1} \\
&& \delta_\varphi \hat{Q}^{+r}_A = \varphi
a^{(rb)}\hat{Q}^{+}_{b\,A}~, \;\; \delta_\varphi \hat{g}^{+r} = \varphi \vert
\kappa \vert q^{+r}~,\;\;\delta_\varphi q^{+ a} = \varphi \, \vert \kappa \vert
\hat{g}^{+ a}  \label{phi1}   \eeqa
(the gauge superfields $W^{++}$, $V^{++}$ have the same transformation laws
as before).

This form of gauge transformations clearly shows that the
corresponding rigid transformations are  linear combinations of {\it
four} independent mutually commuting one-parameter symmetries which are
enjoyed by the free part of the Lagrangian \p{action22}: (a) $SO(2)$ symmetry
realized on the capital index of $\hat{Q}^{+r}_A$; (b) a diagonal $U(1)$
subgroup in the product of two commuting  $SU(2)_{PG}$ groups realized on
$q^{+a}$ and $\hat{g}^{+r}$, with $c^{ik}$ as  the $U(1)$ generator; (c)
$U(1)$ subgroup of the $SU(2)_{PG}$ group acting on $\hat{Q}^{+r}_A$, with
$a^{rs}$ as the $U(1)$ generator; (d) a hyperbolic rotation of $q^{+a}$ and
$\hat{g}^{+r}$,
\beq
\delta\hat{g}^{+r} = \varphi \vert \kappa \vert q^{+r}~,\;\;\delta q^{+ a}
= \varphi \, \vert \kappa \vert \hat{g}^{+ a}~. \label{shiftQ}
\eeq
Note that the bilinear form invariant under \p{shiftQ} is just
$c^{(ij)}(\hat{g}^+_i\hat{g}^+_j - q^+_i q^+_j)$. This explains the
presence of this expression in the $\varepsilon$-Killing potential (first
square  brackets  in \p{action22}):  the $q^+$ term which is needed for making
one of two basic constraints of the theory meaningful and  solvable (see
below)  should be accompanied by the proper $\hat g^+$ term in order to comply
with the symmetry \p{shiftQ}. One is led to $\varepsilon$-gauge the
diagonal $U(1)$ subgroup in the product of two independent $SU(2)_{PG}$ groups
realized on $q^{+a}$ and $\hat{g}^{+r}$ just in order to gain this expression
in the relevant Killing potential. In the HK  limit $\vert\kappa\vert
q^{+a}\rightarrow  u^{+a}, \;\kappa \rightarrow 0$  the symmetry \p{shiftQ}
becomes gauging of the familiar shift symmetry of the free hypermultiplet
action:    \beq \delta \hat{g}^{+r} = \varphi u^{+r}~, \;\;\delta u^{+a} = 0~.
\label{shiftK} \eeq

Thus we come to the conclusion that our original Lagrangian \p{action2} is
the  simplest and natural choice yielding the double Taub-NUT HK action in
the $\kappa \rightarrow 0 $ limit, but it is by no means the unique one.
Indeed,  one could gauge two most general independent combinations of the
four commuting  $U(1)$ symmetries just mentioned. The corresponding
generalization of \p{action22} which still has a smooth $\kappa \rightarrow
0$ limit is as follows
\beqa
{\cal L}^{+4}_{dTN}{}' &=& -q^+_a{\cal
D}^{++}q^{+a} + \hat{Q}^{+}_{rA}{\cal D}^{++}\hat{Q}^{+r}_{A} +
\hat{g}^+_r{\cal D}^{++}\hat{g}^{+r} \nn
&& + \, W^{++}\left[
\hat{Q}^{+a}_A\hat{Q}^+_{a B}\epsilon_{AB} - \kappa^2
c^{(ij)}\left(\hat{g}^+_i\hat{g}^+_j - q^{+}_iq^+_j \right) - \beta_0\,
a^{(rf)}\hat{Q}^+_{rA}\hat{Q}^+_{fA}\right] \nn
&& + \,V^{++}\left[ 2\vert
\kappa \vert \left(q^+\cdot \hat{g}^+\right) -
a^{(rf)}\hat{Q}^+_{rA}\hat{Q}^+_{fA} - \alpha_0\, \kappa^2
c^{(ij)}\left(\hat{g}^+_i\hat{g}^+_j - q^{+}_iq^+_j \right)\right],
\label{actionGen}
\eeqa
with $\alpha_0$ and $\beta_0$ ($[\alpha_0] = -1$, $[\beta_0] = 1$) being two
new real independent parameters. It is straightforward to find the precise
modification of the gauge transformation rules \p{eps1}, \p{phi1}:
\beqa
&&\tilde{\delta}_\varepsilon \hat{Q}^{+r}_A = \delta_\varepsilon
\hat{Q}^{+r}_A + \varepsilon\,\beta_0\, a^{(rb)}\hat{Q}^{+}_{b\,A}~,
\;\;\tilde{\delta}_\varphi
\hat{g}^{+r} = \delta_\varphi\hat{g}^{+r} + \varphi\,\alpha_0 \,\kappa^2
c^{(rn)} \hat{g}^+_n~, \nn
&&\tilde{\delta}_\varphi q^{+ a} = \delta_\varphi
q^{+ a} +  \varphi \, \alpha_0\,\kappa^2 c^{(ab)}q^+_b
\label{modTran}
\eeqa
(the rest of transformations remains unchanged).
\vspace{0.4cm}

\noindent\underline{\it Limits and truncations}\\

In the HK limit the generalized Lagrangian is  reduced to
\beqa
{\cal L}^{+4}_{dTN}{}'(\kappa \rightarrow 0) &=& \hat{Q}^{+}_{rA}
D^{++}\hat{Q}^{+r}_{A} +
\hat{g}^+_r D^{++}\hat{g}^{+r} \nn
&& + \,W^{++}\left[ \hat{Q}^{+a}_A\hat{Q}^+_{a B}\epsilon_{AB} -\beta_0\,
a^{(rf)}\hat{Q}^+_{rA}\hat{Q}^+_{fA} + c^{(ij)}u^{+}_iu^+_j \right] \nn
&& +\,V^{++}\left[ 2\left(u^+\cdot \hat{g}^+\right) -
a^{(rf)}\hat{Q}^+_{rA}\hat{Q}^+_{fA} + \alpha_0\, c^{(ij)}u^{+}_iu^+_j
\right]~. \label{actionGenHK}
\eeqa
It is easy to see that the $\alpha_0$ term in the second bracket in
\p{actionGenHK} can be removed by the redefinition
\beq
\hat{g}^{+r} \;\Rightarrow \; \hat{g}^{+r} -{1\over 2} \alpha_0\,
c^{ri}u^+_i~, \label{Redef}
\eeq
which does not affect the kinetic term of $\hat{g}^{+r}$. At the same time,
no such a redefinition is possible in the QK Lagrangian  \p{actionGen}, so
$\alpha_0$ is the essentially new parameter of the corresponding QK metric.
This $\alpha_0$-freedom disappears in the HK limit.

Thus the associated class of QK metrics includes two extra free parameters
$\alpha_0$ and $\beta_0$ besides the $SU(2)$ breaking
parameters and Einstein constant which characterize the minimal case treated
before. But only one of them, $\beta_0$, is retained in the HK limit. Here we
encounter a new (to the best of our knowledge) phenomenon of violation of the
one-to-one correspondence between the HK manifolds and their QK counterparts.

It remains to understand the meaning of the parameter $\beta_0$. At
$\beta_0 = 0$, we have the $\alpha_0$-modified QK double Taub-NUT action. To
see what happens at non-zero $\beta_0$, it is instructive to take a modified
EH limit in \p{actionGenHK}. Let us redefine
$$
a^{ik} = {1\over
\beta_0}\,\tilde{a}^{ik}
$$
and then put $\hat{g}^{+r} = 0$, $\beta_0 \rightarrow \infty$ with keeping
$\tilde{a}^{ik}$ finite and non-vanishing. Then \p{actionGenHK} goes into
\beqa
{\cal L}^{+4}_{EH}{}'(\kappa \rightarrow 0) =
\hat{Q}^{+}_{rA}D^{++}\hat{Q}^{+r}_{A} +   W^{++}\left[
\hat{Q}^{+a}_A\hat{Q}^+_{a B}\epsilon_{AB}
-\tilde{a}^{(rf)}\hat{Q}^+_{rA}\hat{Q}^+_{fA} + c^{(ij)}u^{+}_iu^+_j \right].
\label{actionGenHK1}
\eeqa
It is shown in section 5.5 that this HSS action produces a generalization of
the standard two-centre Eguchi-Hanson metric by bringing in two unequal
``masses'' $1-a$ and $1+a$ in the numerators of poles in the relevant
two-centre potential, with $a = \sqrt{{1\over
2}\tilde{a}^{ik}\tilde{a}_{ik}}$ ($c^{ik}$ specifies the centres like in the
standard EH case \cite{giot}). Then it is clear that the action \p{actionGenHK}
describes a similar non-equal masses modification of the double Taub-NUT
metric as a non-trivial ``hybrid'' of the Taub-NUT and unequal masses EH
metrics, with $\beta_0$ measuring the ratio of the masses.

The general Lagrangian \p{actionGen} has still two commuting rigid $U(1)$
symmetries which constitute the $U(1)\times U(1)$ isometry of the related QK
metric. As distinct from the QK Taub-NUT and EH truncations \p{TN} and \p{EH}
of \p{dTN1}, in which the isometries are enhanced to $U(2)$ \cite{bgio,IvV},
the same truncations made in the Lagrangian \p{actionGen} lead to generalized
QK Taub-NUT and EH metrics having only $U(1)\times U(1)$ isometries. In the
QK Taub-NUT truncation which is performed by putting $\hat{Q}^{+a}_2 =
\beta_0 = 0~, c^{ik} = 0~, \alpha_0 c^{ik} \equiv \tilde{c}^{ik} \neq 0$ in
\p{actionGen},
\beqa
{\cal L}^{+4}_{dTN}{}' &\Rightarrow & {\cal L}^{+4}_{TN}{}' = -q^+_a{\cal
D}^{++}q^{+a} + \hat{Q}^{+}_{r1}{\cal D}^{++}\hat{Q}^{+r}_{1} +
\hat{g}^+_r{\cal D}^{++}\hat{g}^{+r} \nn
&& + \,\,V^{++}\left[ 2\vert
\kappa \vert \left(q^+\cdot \hat{g}^+\right) -
a^{(rf)}\hat{Q}^+_{r1}\hat{Q}^+_{f1} - \kappa^2
\tilde{c}^{(ij)}\left(\hat{g}^+_i\hat{g}^+_j - q^{+}_iq^+_j \right)\right]~,
\label{GenTN}
\eeqa
this isometry is again enhanced to $U(2)$ after taking the HK limit, because
any dependence on the breaking parameter $\tilde{c}^{ik}$ disappears in this
limit (after the redefinition like \p{Redef}). At the same time, in the QK EH
truncation ($\hat{g}^{+r} = \alpha_0 = a^{rf} = 0~, \;\beta_0\,a^{rf} \equiv
\tilde{a}^{rf} \neq 0$ in \p{actionGen}) the $U(1)\times U(1)$ isometry is
retained in the HK limit, as clearly seen from the form of the limiting
HK Lagrangian \p{actionGenHK1} (parameters $\tilde{a}^{ik}$ break $SU(2)_{PG}$
and $c^{ik}$ break the $SU(2)$ which rotates harmonics).
\vspace{0.4cm}

\noindent\underline{\it Alternative HSS quotient}\\

Finally, we wish to point out that the QK sigma model
actions we considered up to now give rise to the QK metrics which are
one or another generalization of the HK  double Taub-NUT metric. This is
closely related to the property that one of the symmetries of the free QK
action of $(q^{+a},\hat{Q}^{+a}_A, \hat{g}^{+r})$ which we gauge always
includes as a part the hyperbolic $\hat{g}^{+r}, q^{+a}$ rotation \p{shiftQ}
becoming a pure shift \p{shiftK} of $g^{+r}$ in the HK limit. This ensures the
existence of the QK Taub-NUT truncation for the considered class of QK metrics.
An essentially different class of QK metrics can be constructed by gauging two
independent combinations of those mutually commuting $U(1)$ symmetries
of the free action which are realized as the homogeneous phase transformations
of the involved superfields. The  most general gauged QK sigma  model of this
kind is specified by the following superfield Lagrangian
\beqa
{\cal L}^{+4}_{dEH} &=&
-q^+_a{\cal D}^{++}q^{+a} + \hat{Q}^{+}_{rA}{\cal D}^{++}\hat{Q}^{+r}_{A} +
\hat{g}^+_r{\cal D}^{++}\hat{g}^{+r} \nn
&& + \, W^{++}\left[
\hat{Q}^{+a}_A\hat{Q}^+_{a B}\epsilon_{AB} +
\gamma_0 d^{(ik)}\hat{g}^+_i\hat{g}^+_k +
\beta_0 a^{(rf)}\hat{Q}^+_{rA}\hat{Q}^+_{fA}+ c^{(ik)}\kappa^2q^{+}_iq^+_j
\right] \nn
&& + \,V^{++}\left[d^{(ik)}\hat{g}^+_i\hat{g}^+_k
- a^{(rf)}\hat{Q}^+_{rA}\hat{Q}^+_{fA} + \alpha_0 \kappa^2
c^{(ij)}q^{+}_iq^+_j \right]~,
\label{actionGen2}
\eeqa
where the involved constants are different from those in \p{actionGen},
despite being denoted by the same letters. To see to which kind of the
4-dimensional HK sigma model the QK Lagrangian \p{actionGen2} corresponds, let
us examine  its HK limit
\beqa
{\cal L}^{+4}_{dEH}(\kappa \rightarrow 0) &=&
\hat{Q}^{+}_{rA}{D}^{++}\hat{Q}^{+r}_{A} +
\hat{g}^+_r{D}^{++}\hat{g}^{+r} \nn
&& + \, W^{++}\left[
\hat{Q}^{+a}_A\hat{Q}^+_{a B}\epsilon_{AB} +
\gamma d^{(ik)}\hat{g}^+_i\hat{g}^+_k +
\beta a^{(rf)}\hat{Q}^+_{rA}\hat{Q}^+_{fA}+ c^{(ik)}u^{+}_iu^+_j
\right] \nn
&& + \,V^{++}\left[d^{(ik)}\hat{g}^+_i\hat{g}^+_k
- a^{(rf)}\hat{Q}^+_{rA}\hat{Q}^+_{fA} + \alpha
c^{(ij)}u^{+}_iu^+_j \right]~.
\label{actionHK3}
\eeqa
Under the truncation $\hat{g}^{+ r} =0$, $\alpha_0 = 0$, $\beta_0 a^{rf} \equiv
\tilde{a}^{rf}\neq 0~,\, a^{rf} =0$ it goes into the Lagrangian
\p{actionGenHK1} which corresponds to the EH model with unequal masses, while
under the truncation  $Q^{+a}_2 =0, Q^{+a}_1 \equiv Q^{+a}$,  $\gamma_0 =
\beta_0 = 0$, $\alpha_0 c^{ik} \equiv \tilde{c}^{ik} \neq 0~,\, c^{ik} =0$ it
is reduced to the following expression
\beqa
{\cal L}^{+4}_{EH}{}''= \hat{Q}^{+}_{r}{D}^{++}\hat{Q}^{+r} +
\hat{g}^+_r{D}^{++}\hat{g}^{+r} + \,V^{++}\left[d^{(ik)}\hat{g}^+_i\hat{g}^+_k
- a^{(rf)}\hat{Q}^+_{rA}\hat{Q}^+_{fA} +
\tilde{c}^{(ij)}u^{+}_iu^+_j \right]~.
\label{actionEH2}
\eeqa
This HSS Lagrangian can be shown to yield again a EH sigma model with unequal
masses. The parameters of this model are different from those pertinent to
the first truncation. Thus \p{actionHK3} defines a ``hybrid'' of two different
EH sigma models, and the associated QK sigma model \p{actionGen2} could be
called the ``QK double EH sigma model''. \footnote{We
expect that the related QK metrics fall into the class of QK metrics described
by the Plebanski-Demianski Ansatz \cite{pd}; this is not the case for the QK
double Taub-NUT metrics, see section 6.4.}.

As the final remark, we note that in the system of three hypermultiplets in
the HK case one can define mutually-commuting independent shifting symmetries
of the form \p{shiftK} separately for each hypermultiplet. Accordingly,
one can use them to define different HSS quotient actions (actually, all
such actions, with at least two independent shifting symmetries \p{shiftK}
gauged, yield the Taub-NUT sigma model, while those where all three such
symmetries
are gauged yield a trivial free 4-dimensional HK sigma model). No such an
option exists in the QK case: any other hyperbolic rotation like \p{shiftQ}
(in the planes $(\hat{Q}^{+a}_1, q^{+i})$ or $(\hat{Q}^{+a}_2, q^{+i})$) does
not commute with \p{shiftQ} and the third one of the same kind. For this
reason, we are allowed to use only one such hyperbolic symmetry in the gauged
combinations of independent $U(1)$ symmetries in the course of constructing
the relevant HSS quotient actions. Of course, this is related to the fact that
the full symmetry of the ``flat'' QK action of $(q^{+a}, \hat{Q}^{+a}_A,
\hat{g}^{+r})$ is $Sp(1,3)$, while the analogous symmetry of
the relevant limiting HK action is a contraction  of $Sp(1,3)$, with a bigger
number of the mutually commuting abelian subgroups.

\setcounter{equation}{0}
\section{From the HSS actions to QK metrics}

\subsection{Preparatory steps}
As already mentioned, the basic advantage of the HSS quotient as compared
to the approach based on solving nonlinear harmonic equations is the
opportunity  to choose the WZ gauge for $W^{++}$ and $V^{++}$ by using the
$\varepsilon$ and $\varphi$ gauge freedom (see \p{eps}, \p{phi}). In this
gauge the harmonic differential equations for the lowest components
$f^{+a}(x,u), \hat{F}^{+r}_A(x,u)$ and  $\hat{g}^{+ r}(x,u)$ of the
superfields $q^{+a}(\zeta), \hat{Q}^{+ r}_A(\zeta)$ and $\hat{g}^{+ r}(\zeta)$
become linear and can be straightforwardly solved.

In the WZ gauge the gauge superfields has the following short expansion
\beqa
&& W^{++}(\zeta) = i\theta^+\sigma^m\bar\theta^+ W_m(x) +
(\theta^+)^2(\bar\theta^+)^2 P^{(ik)}(x)u^-_iu^-_k~, \nn
&& V^{++}(\zeta) = i\theta^+\sigma^m\bar\theta^+ V_m(x) +
(\theta^+)^2(\bar\theta^+)^2 T^{(ik)}(x)u^-_iu^-_k \label{WZ}
\eeqa
(like in \p{thetaexp1}, we omitted possible terms proportional to the monomials
$(\theta^+)^2$ and $(\bar\theta^+)^2$ because the equations of motion for the
corresponding fields are irrelevant to our problem of computing the final
target QK metrics). At the intermediate steps it is convenient to deal with the
hypermultiplet superfields $\hat{Q}^{+a}_A, \hat{g}^{+r}$ related to the
original superfields by \p{hat}. They have the same $\theta$ expansions
\p{thetaexp1}, with ``hat'' above all the component fields. Due to the
structure of the WZ-gauge \p{WZ}, the highest components in the
$\theta$ expansions of the superfields $\hat{Q}^{+a}_A, \hat{g}^{+r}$ and
$q^{+a}$ ($\hat{G}^{-3 a}_A(x,u)$, $\hat{g}^{-3r}(x,u)$ and
$f^{-3a}(x,u)$) appear only in the  kinetic part of \p{action2}. This
results in the linear harmonic equations for $f^{+a}(x,u), \hat{F}^{+r}_A(x,u)$
and $\hat{g}^{+r}(x,u)$:
\beqa   &&\partial^{++}f^{+a} = 0\;
\Rightarrow \;f^{+a} = f^{ai}(x)u^+_i~, \;  \partial^{++}\hat{F}^{+r} = 0\;
\Rightarrow \;\hat{F}^{+b}_A = \hat{F}^{bi}_A(x)u^+_i~, \nn  &&
\partial^{++}\hat{g}^{+r}= 0\; \Rightarrow \;\hat{g}^{+r} =
\hat{g}^{ri}(x)u^+_i~. \label{solu}
\eeqa
It is easy to check that these equations are covariant under the $SU(2)_{loc}$
transformations \p{fFSU2} which act on $f^{+a}, \hat{F}^{+r}_A = \vert \kappa
\vert (u^-\cdot f^+) F^{+r}_A$ and $\hat{g}^{+r}= \vert
\kappa \vert (u^-\cdot f^+) g^{+r}$ as  follows:
\beqa
&&\delta^* f^{+a} = \lambda^{+-}f^{+a} - \lambda^{++}\partial^{--}f^{+a}~,
\quad  \delta^* \hat{F}^{+r}_A = \lambda^{+-}\hat{F}^{+r}_A -
\lambda^{++}\partial^{--}\hat{F}^{+r}_A~, \nn
&& \delta^* \hat{g}^{+r} = \lambda^{+-}\hat{g}^{+r} -
\lambda^{++}\partial^{--}\hat{g}^{+r}
 \label{fFgSU2}
\eeqa
(in checking this, one must use the properties $\partial^{++}\lambda^{++} =
0~, \partial^{++}\lambda^{+-} = \lambda^{++}$, $[\partial^{++},
\partial^{--}] = \partial^0 \equiv u^{+i}\partial/\partial u^{+i} -
u^{-i}\partial/\partial u^{-i}$ and $\partial^0 (f^{+a}, \hat{F}^{+r},
\hat{g}^{+r} ) =  (f^{+a}, \hat{F}^{+r}, \hat{g}^{+r} )$). These
transformations entail the following ones for the bosonic fields of physical
dimension  \beq
\delta f^{ai}(x) = \lambda^i_{\;k}(x)f^{ak}(x)~, \quad \delta
\hat{F}^{ri}_A(x) = \lambda^i_{\;k}(x)\hat{F}^{rk}_A(x)~, \quad \delta
\hat{g}^{ri}(x) = \lambda^i_{\;k}(x)\hat{g}^{rk}(x)~. \label{transffFg}
\eeq

This step is common for all QK sigma model actions considered in the previous
section. The next common step is to vary with respect to the SG fields $D(x)$
and ${\cal V}^{ik}_m(x)$ in order to obtain the appropriate particular forms of
the constraint \p{constr} and the expression \p{exprV}. Bearing in mind
the harmonic ``shortness'' \p{solu}, we find
\beqa
&&{\kappa^2\over 2}f^2 = 1 +{\kappa^2 \over 2}(\hat{F}^2 +\hat{g}^2)~,
\label{constrPart} \\
&& {\cal V}^{ik}_m = \kappa^2 \left(f^{a(i}\partial_m
f_a^{j)} -  \hat{F}^{r(i}_A\partial_m \hat{F}^{j)}_{r A} -
\hat{g}^{r(i}\partial_m \hat{g}^{j)}_r \right)~, \label{composV}
\eeqa
where
$$
f^2 = f^{ai}f_{ai}~, \quad \hat{F}^2 = \hat{F}^{ri}_A\hat{F}_{ri\,A}~,
\quad \hat{g}^{ri}\hat{g}_{ri}~.
$$
Taking into account the constraint \p{constrPart}, it is easy to check that
the $SU(2)_{loc}$ transformation laws \p{fFgSU2} imply just the
transformation law \p{VDSU2} for the composite gauge field \p{composV} .

One more common step is enforcing the gauge \p{gauge}
\beq
f^i_a(x) = \delta^i_a\, \omega(x) \;\Rightarrow \; f^{ai}f_{ai} \;\Rightarrow
\; 2 \omega^2~. \label{gauge2}
\eeq
For what follows it will be useful to give how the residual gauge symmetries
of the WZ gauge \p{WZ} with the parameters $\varepsilon(x)=
\varepsilon(\zeta)\vert$ and $\varphi(x) = \varphi(\zeta) \vert $ are
realized in the gauge \p{gauge2} (in the general case of gauge
symmetries \p{modTran}, \p{eps1}, \p{phi1})
\beqa
&&\tilde{\delta}_\varepsilon \hat{F}^{ri}_A = \varepsilon \epsilon_{AB}
\hat{F}^{ri}_B + \varepsilon \,\beta_0 a^{rs}\hat{F}_{s\,A}^{\;i} -
\lambda^{ik}_\varepsilon \hat{F}^{r}_{\;k\,A}~, \quad
\tilde{\delta}_\varepsilon \hat{g}^{ri} = \varepsilon\,\kappa^2
c^{rn}\hat{g}_{n}^{\,i} - \lambda^{ik}_\varepsilon \hat{g}^{r}_{\;k}~,
\label{eps22} \\
&& \tilde{\delta}_\varphi
\hat{F}^{ri}_A =  \varphi \,a^{rs}\hat{F}^{\;i}_{s\,A} - \lambda^{ik}_\varphi
\hat{F}^{r}_{\;k\,A}~, \quad \tilde{\delta}_\varphi \hat{g}^{ri} =
\varphi\,\vert \kappa \vert \epsilon^{ri}\omega +\varphi \,\alpha_0 \,\kappa^2
c^{rs}\hat{g}_s^{\;i} - \lambda^{ik}_\varphi \hat{g}^{r}_{\;k}~, \nn
&& \tilde{\delta}_\varphi \omega = {1\over 2} \varphi\,\vert \kappa \vert
(\epsilon_{ia}\hat{g}^{ai})~, \label{phi22}
\eeqa
where $\lambda^{ik}_\varepsilon $, $\lambda^{ik}_\varphi $ are the
parameters of two different induced $SU(2)_{loc}$ transformations needed to
preserve  the gauge \p{gauge2}
\beq
\lambda^{ik}_\varepsilon = -\varepsilon\,\kappa^2\,c^{ik}~, \quad
\lambda^{ri}_\varphi  =  - \varphi \left( \frac{\vert \kappa
\vert}{\omega} \,\hat{g}^{(ri)} + \alpha_0\, \kappa^2 c^{ri}
\right)~. \label{induced}
\eeq
From now on, we fully fix the residual $\varphi(x)$ gauge symmetry by
gauging  away the singlet part of $g^{ri}(x)$:
\beq
\epsilon_{ir}g^{ri}(x) = 0 \;\Rightarrow \; g^{ri}(x) = g^{(ri)}(x)~.
\label{gaugeg}
\eeq
The residual $SO(2)$ gauge freedom, with the parameter $\varepsilon(x)$, will
be kept for the moment.

We shall explain further steps on the example of the simplest QK
double Taub-NUT action \p{dTN1}, \p{action2} and then indicate
the modifications which should be made in the resulting physical bosons
action in order to encompass the general case \p{actionGen}.

These steps are technical (though sometimes amounting to rather lengthy
computations) and quite similar to those expounded in \cite{IvV} on the
examples  of the QK extensions of the Taub-NUT and EH metrics. So here we
shall describe them rather schematically.

Firstly one substitutes the solution \p{solu} back into the action \p{dTN1},
\p{action2} (with the  $\theta$-integration performed) and varies with respect
to the remaining non-propagating (vector) fields of the hypermultiplet
superfields ($A^{-a}_m(x,u)$, $\hat{B}^{-a}_{Am}(x,u)$ and
$\hat{b}^{-r}_m(x,u)$ in the $\theta$-expansions of $q^{+a}$, $\hat{Q}^{+a}_A$
and $\hat{g}^{+r}$, respectively). Then  one  substitutes the resulting
expressions for these fields into the action (together with those for
$\omega(x) $ and ${\cal V}^{ik}_m(x)$, eqs. \p{constrPart}, \p{exprV}) and
performs the $u$-integration. At this stage it is convenient to redefine the
remaining fields as follows \footnote{This relation was misprinted in
\cite{civ}.}
\beq
F^{ai}_A = {1\over \kappa \omega}\hat{F}^{ai}_A~, \quad g^{ri} =
{2\over \kappa \omega} \hat{g}^{ri}~.
\eeq
In terms of the redefined fields and with taking account of the gauges
\p{gauge2}, \p{gaugeg}, the composite fields $\omega$ and ${\cal V}^{ij}_m$
  are given by the following expressions:
\beq
\kappa\, \omega = {1\over \sqrt{1 - {\lambda\over 2} g^2- 2\lambda F^2
}}~, \quad {\cal V}^{(ij)}_m = - 16 \lambda^2\omega^2 \left[
F^{a(i}_A\partial_m F^{\,j)}_{a A} + {1\over 4} g^{r(i}\partial_m g^{j)}_r
\right]~,   \eeq
where
\beq
F^2 \equiv F^{ai}_AF_{ai A}~, \quad g^2 \equiv g^{ri}g_{ri}~, \quad \lambda
\equiv {\kappa^2\over 4}~.
\eeq

After substituting everything back into the action we get the following
intermediate expression for the $x$-space Lagrangian density ${\cal
L}_{dTN}(x)$:
\beq
{\cal L}_{dTN}(x) = {\cal L}_0(x) + {\cal L}_{vec}(x)~,
\eeq
where
\beq
{\cal L}_0(x) = \frac 1{{\cal D}^2}\left\{{\cal D}\left(X+\frac Y4\right)+
\la\left(g^2\cdot\frac Y8+2T\right)\right\} \label{dist0}
\eeq
with
\beqa
&& {\cal D}=1-\frac{\la}{2}\,g^2-2\la\,F\,^2,\;
\;X=\frac{1}{2}\,\partial^mF_{ai\,A}\, \partial_m F^{ai}_{A},\quad
Y=\frac{1}{2}\, \partial^m g_{ij}\, \partial_m g^{ij},\nonumber \\
&& T=F^{\,i}_{a\,B}\,\partial^m
F^{aj}_{B}\left(F_{ai\,A}\,\partial_m F^{a}_{\,j\,A}
+\frac{1}{2}\,g_{ir}\,\partial_m g^r_{~j}\right)~,\label{defD}
\eeqa
and
\beq
{\cal L}_{vec}(x) = {1\over {\cal D}} \left[ \alpha (W^mW_m) + \beta (V^mV_m) +
\gamma (W^mV_m) + W^m K_m + V^mJ_m \right]~,
\eeq
with
\beqa
&& J_m=\frac 12\ a^{ab}\ F_{a\,A}^{\,i}\ \partial_mF_{bi\,A},\qq
K_m=-\frac 12\,\eps_{AB}\ F^{ai}_{A}\,\partial_mF_{ai\,B}
-\frac{\la}{2}\ c_{ij}\ g^i_{~s}\ \partial_mg^{sj}, \nn
&& \alf=\frac12\left(\frac{F\,^2}{4}-\la\,\hat{c}^2
+\frac{\la^2}{2}\,\hat{c}^2\,g^2\right),\quad
\be=\frac14\left(1+\frac{\hat{a}^2}{4}\,F\,^2 -
\frac{\la}{2}\,g^2\right),\nn
&& \ga=\frac 14\,a^{ab}\,F^{\,i}_{a\,A}\,F_{bi\,B}\,\eps_{AB}-\la(c\cdot g)~,
\label{def1} \\
&& \hat{c}^2 \equiv c^{ik}c_{ik}~, \quad \hat{a}^2 = a^{ab}a_{ab}~.
\label{squar1}
\eeqa
After integrating out the non-propagating gauge fields $W^m(x)$ and $V^m(x)$,
the part ${\cal L}_{vec}$ acquires the typical nonlinear sigma model form
\beq
{\cal L}_{vec}\;\Rightarrow \; \frac{1}{{\cal D}}\,Z~, \quad Z =
\frac{1}{4\,\alf\,\be-\ga^2}\,  \left\{\ga\, (J\cdot K)-\alf\, (J\cdot J)
-\be\, (K\cdot K)\right\}~. \label{vec0}
\eeq

The resulting sigma model action should be supplemented by two algebraic
constraints on the involved fields
\beqa
&& F^{a(i}_A\,F^{\,j)}_{a\,B}\,\eps_{AB} -\la\,g^{(li)}\,
g^{(rj)}\,c_{(lr)}+c^{(ij)}=0~,  \label{contr1} \\
&& g^{ij}- a^{ab}\,F^{\,i}_{a\,B}\,F^{\,j}_{b\,B} =0~, \label{contr2}
\eeqa
which follow from varying the action with respect to the auxiliary fields
$P^{(ik)}(x)$ and $T^{(ik)}(x)$ in the WZ gauge \p{WZ}.
Keeping in mind these 6 constraints and one residual gauge ($SO(2)$)
invariance, one is left just with four independent bosonic target coordinates
as compared with eleven such coordinates explicitly present in \p{dist0},
\p{vec0}. The problem now is to solve eqs. \p{contr1}, \p{contr2}, and thus
to obtain the final sigma model action with 4-dimensional QK target manifold.
This will be the subject of our further presentation.

Here, as the convenient starting point for the geometrical treatment in section
5, it is worth to give how the full distance looks before solving the
constraints \p{contr1}, \p{contr2}
\beq
\EuFrak{g} = \frac{1}{{\cal D}^2}\left\{{\cal
D}\left(X'+ Z'+\frac{Y'}{4}\right)+
\la\left(g^2\cdot\frac{Y'}{8}+2\,T'\right)\right\}~. \label{dist00}
\eeq
The quantities with ``prime'' are obtained from those defined above by
replacing altogether ``$\partial_m$'' by ``$d$'', thus passing to the
distance  in the target space. For instance,
\beq
X'=\frac{1}{2}\,dF_{ai\,A}\, dF^{ai}_{A},\quad Y'=\frac{1}{2}\, dg_{ij}\,
dg^{ij}~. \label{dist1}
\eeq
Note that this metric includes three free parameters. These are the Einstein
constant related to $\,\la$ ($\la \equiv {\kappa^2\over 4}$), and two $SU(2)$
breaking parameters: the triplet $\,c^{(ij)},$ which breaks the $\,SU(2)_{\rm
SUSY}\,$ to $\,U(1),$ and the triplet  $\,a^{(ab)},$ which breaks the
Pauli-G\"ursey $\,SU(2)\,$ to $\,U(1)$. The final isometry group is therefore
$\,U(1)\times U(1)$. Constraints \p{contr1}, \p{contr2} are manifestly
covariant under these isometries. For convenience, from now on we choose the
following frame with respect to the broken $SU(2)$ groups
\beq
c^{12}=ic,\quad c^{11}=c^{22}=0,\qq
a^{12}=ia,\quad a^{11}=a^{22}=0~, \label{frame}
\eeq
with real parameters $\,a\,$ and $\,c$. In this frame, the squares \p{squar1}
become
$$
\hat{c}^2 = 2 c^2~, \quad \hat{a}^2 = 2 a^2~.
$$

Let us now discuss which modifications the distance \p{dist00}
undergoes if one starts from the general QK double Taub-NUT action
corresponding to the Lagrangian \p{actionGen}. Since the difference between
\p{dTN1} and \p{actionGen} is solely  in the structure of supercurrents
(Killing potentials) to which gauge superfields $W^{++}$ and $V^{++}$ couple,
the only modifications entailed by passing to \p{actionGen} are the
appropriate changes in the   $Z'$-part of \p{dist00} and in  the constraints
\p{contr1}, \p{contr2}.  Namely, one should  make the following replacements
in $Z'$:  \beqa
\alpha \;&\Rightarrow &
\; \hat{\alpha} =  \alpha + {1\over 16}\,\beta_0^2\, \hat{a}^2\, F^2 + {1\over
4} \beta_0\,a^{rf}F^i_{rA}F_{fi\,B}\epsilon_{AB}~, \nn
\beta \;&\Rightarrow & \; \hat{\beta} = \beta - \lambda\, \alpha_0\, (g\cdot
c) - {1\over 2}\, \alpha_0^2\, \hat{c}^2 \left(1 -{\lambda \over 2}\,
g^2\right), \nn
\gamma \;&\Rightarrow &\; \hat{\gamma} = \gamma + {1\over
8}\,\beta_0\, \hat{a}^2\, F^2 -\lambda\, \alpha_0\,\hat{c}^2 \left(1 -{\lambda
\over 2}\, g^2\right), \nn
K_m \;&\Rightarrow &\; \hat{K}_m = K_m + {1\over 2}\,\beta_0\,
a^{rf}F^i_{rA}d F_{fi\,A}~, \nn
J_m \;&\Rightarrow &\; \hat{J}_m = J_m -
{1\over 2}\,\lambda\,\alpha_0\, c^{lr}g_{lk}\, dg ^k_{\;\;r}
\eeqa
and pass to the following modification of the constraints \p{contr1},
\p{contr2}:
\beqa
&& F^{a(i}_A\,F^{\,j)}_{a\,B}\,\eps_{AB} -\la\,g^{(li)}\,
g^{(rj)}\,c_{(lr)} -\beta_0\,
a^{ab}\,F^{\,i}_{a\,B}\,F^{\,j}_{b\,B}+c^{(ij)}=0~,  \label{constr1a} \\  &&
g^{ij}-a^{ab}\,F^{\,i}_{a\,B}\,F^{\,j}_{b\,B} +\alpha_0\left[ c^{ij}
- \la\,g^{(li)}\, g^{(rj)}\,c_{(lr)}\right] = 0~. \label{constr2a}
\eeqa

\subsection{Solving the constraints}
In order to find the final form of the QK target metric corresponding to the
HSS Lagrangian \p{action2} or its generalization \p{actionGen}, we should
solve the constraints \p{contr1}, \p{contr2} or their generalization
\p{constr1a}, \p{constr2a}. It is a non-trivial step to find the true
coordinates to solve these constraints. Indeed, a direct substitution of
$g^{ij}$ from \p{contr2} into \p{contr1} gives a quartic constraint for
$F^{ai}_A$ which is very difficult to solve as compared to the HK case
\cite{giot,gorv}  where the analogous constraint is merely quadratic. In the
general case \p{constr1a}, \p{constr2a} the situation is even worse.

In view of these difficulties, it proves more fruitful to take as
independent  coordinates just the components of the triplet $g^{(ri)}$,
$$
g^{12}= g^{21}\equiv iah~,\quad \ol{h}=h~,\qq g^{11} \equiv g~,\quad
g^{22}=\ol{g}~,
$$
and one angular variable from $F^{ai}_A~.$ Then the above 6 constraints and
one residual gauge invariance  (the $\varepsilon(x)$ one)
allow us to eliminate the remaining 7 components of $F^{ai}_A$ in terms of
4 independent coordinates thus defined. Following the same strategy as in
the previous subsection, we shall first explain how to solve eqs.
\p{contr1} and \p{contr2} in this way and then indicate the modifications
giving rise to the solution of the general two-parameter set of constraints
\p{constr1a}, \p{constr2a}.
We relabel the  components of $F^{ai}_A$ as follows
$$
\left\{\barr{l}
\dst F^{a=1\ i=2}_{A=1}=\frac 12({\cal F}+{\cal K})~,\quad  F^{a=1\
i=1}_{A=1}=\frac 12({\cal P}+{\cal V})~,\\[5mm] \dst F^{a=1\ i=2}_{A=2}=\frac
1{2i}({\cal F}-{\cal K})~,\quad  F^{a=1\ i=1}_{A=2}=\frac 1{2i}({\cal P}-{\cal
V})~,\\[5mm] \quad F^{a=2\ i=1}_{A}=-\ol{F^{a=1\ i=2}_{A}}~, \quad F^{a=2\
i=2}_{A}=\ol{F^{a=1\ i=1}_{A}}~,\earr\right.
$$
and substitute this into \p{contr1}, \p{contr2}. After some simple algebra,
the constraints can be equivalently rewritten in the following form
\beqa
&&\mbox{(a)}\;\;{\cal P}\bar{\cal F} = - {i\over 2a} A_-~, \;\;\mbox{(and
c.c.)}~; \quad \mbox{(b)}\;\; {\cal V}\bar{\cal K} =  - {i\over 2a} A_+~,\;\;
\mbox{(and c.c.)}~; \label{constrab} \\
&& \mbox{(c)}\;\; {\cal F}\bar{\cal F} - {\cal P}\bar{\cal P} = B_+~; \quad
\mbox{(d)}\;\; {\cal V}\bar{\cal V} - {\cal K}\bar{\cal K} = B_-~.
\label{constrcd}
\eeqa
Here
$$
A_{\pm}=1\pm 2\la a^2c\,h~, \quad B_{\pm}=c(1+\la a^2 r^2)\pm
h\,A_{\mp}~, \quad
r^2=h^2+t^2~, \quad g\bar g = a^2 t^2~.
$$
Next, one expresses $\bar{\cal P}$ and $\bar{\cal K}$ from \p{constrab} and
substitutes them into \p{constrcd}, which gives two quadratic equations
for ${\cal F}\bar{\cal F} \equiv X$ and ${\cal V}\bar{\cal V} \equiv Y$,
\beq
X^2 - X\,B_+ - {1\over 4}\, t^2 A^2_- = 0~, \quad  Y^2 - Y\,B_- - {1\over
4}\, t^2 A^2_+ = 0~.
\eeq

Solving these equations, selecting the solution which is regular in the limit
$g = \bar g = h =0$ and properly fixing the phases of ${\cal F}, {\cal P},
{\cal V}$ and ${\cal K}$ in terms of the phase of $g$ with taking account
of the residual $\varepsilon(x)$ gauge freedom, we find the general solution
of  \p{contr1}, \p{contr2} in the following concise form
\beqa
&& \dst {\cal P}=-iM\,e^{i(\phi+\alf/\rho_-
-\mu\rho_+)}~, \qq\qq  \dst {\cal F}=R\,e^{i(\phi-\mu\rho_-)}~, \nn
&& \dst {\cal K}=iS\,e^{i(\phi-\alf/\rho_- +\mu\rho_+)}~,  \qq\qq\quad \;
\dst {\cal V}=L\,e^{i(\phi+\mu\rho_-)}~,  \nn
&& \rho_{\pm}=1\pm 4\la c  \label{sol1}
\eeqa
and
\beq
g=at\,e^{i(\alf/\rho_- - 8\mu \la c )}~. \label{sol2}
\eeq
The various functions involved are
$$\barr{ll}
\dst L=\sqrt{\frac 12(\sqrt{\De_-}+B_-)}~, & \qq\qq
\dst R=\sqrt{\frac 12(\sqrt{\De_+}+B_+)}~,\\[4mm]
\dst M=\sqrt{\frac 12(\sqrt{\De_+}-B_+)}~, & \qq\qq
\dst S=\sqrt{\frac 12(\sqrt{\De_-}-B_-)}~,\earr$$
where
$$
\Delta_{\pm} = B^2_{\pm} + t^2 A^2_{\mp}~.
$$

The true coordinates are $\,(\phi,\,\alf,\,h,\,t)$. An extra angle $\mu$
parametrizes the residual local $SO(2)$ transformations which act as shifts
of $\mu$ by the parameter $\varepsilon(x)$, $\mu \rightarrow \mu
+\varepsilon~.$ To see this, one must rewrite the
$\varepsilon$-transformation law of $F^{ri}_A$ following from that of
$\hat{F}^{ri}_A$, eq. \p{eps22} (at $\beta_0 = \alpha_0 =0$),
$$\delta_\varepsilon F^{ri}_A = \varepsilon\, \epsilon_{AB}F^{ri}_B
+\varepsilon\,\kappa^2 c^{ik}F^{r}_{\;k\,A}~, $$ in terms of the newly
defined variables and in the $SU(2)$ frame \p{frame}
\beqa
&& \delta_\varepsilon {\cal F} = -i\varepsilon\,\rho_- {\cal F}~, \quad
\delta_\varepsilon {\cal V} = i\varepsilon\,\rho_- {\cal V}~, \quad
\delta_\varepsilon {\cal P} = -i\varepsilon\,\rho_+ {\cal P}~, \quad
\delta_\varepsilon {\cal K} = i\varepsilon\,\rho_+ {\cal K}~, \nn
&& \delta_\varepsilon  h =0~, \quad \delta_\varepsilon g = -8i \varepsilon\,
\lambda\, c\,g~. \label{tranFVKP}
\eeqa
As a consequence of gauge invariance of \p{dist0}, the final form of the
metric should not depend on $\mu $ and we can choose the latter at will. For
instance, we can change the precise dependence of phases in \p{sol1},
\p{sol2} on $\phi$ and $\alpha~.$ In what follows we shall stick just to the
above parametrization. Explicitly keeping $\mu $ at the intermediate steps of
 calculations is a good self-consistency check: this gauge parameter should
fully drop out from the correct final expression for the metric.

Finally, let us indicate the modifications which should be made in the above
solution  to adapt it to the general set of constraints \p{constr1a},
\p{constr2a}. It is convenient to represent the latter in the following
equivalent form
\beqa  && F^{a(i}_A\,F^{\,j)}_{a\,B}\,\eps_{AB} -
\beta_0 \,g^{(ij)}  +(1 -\alpha_0\beta_0)\left[c^{(ij)} - \la\,g^{(li)}\,
g^{(rj)}\,c_{(lr)} \right]=0~,  \label{constr11a} \\   &&
g^{ij}-a^{ab}\,F^{\,i}_{a\,B}\,F^{\,j}_{b\,B} +\alpha_0\left[c^{ij} -
\la\,g^{(li)}\, g^{(rj)}\,c_{(lr)} \right] = 0~. \label{constr22a}
\eeqa
Then, following the same line as in the case of $\beta_0 = \alpha_0 = 0$, one
gets the general solution in the form
\beqa  && \dst {\cal
P}=-i\tilde{M}\,e^{i(\phi+\alf/\rho_- -\mu\rho_+)}e^{-i\mu\beta_0 a}~, \qq\qq
\dst {\cal F}=\tilde{R}\,e^{i(\phi-\mu\rho_-)}e^{-i\mu\beta_0 a}~, \nn && \dst
{\cal K}=i\tilde{S}\,e^{i(\phi-\alf/\rho_- +\mu\rho_+)}e^{-i\mu\beta_0 a}~,
\qq\qq\quad \; \dst {\cal V}=\tilde{L}\,e^{i(\phi+\mu\rho_-)}e^{-i\mu\beta_0
a}~, \label{sol3}
\eeqa
where the functions with ``tilde'' are related to those defined earlier by the
following replacements
\beqa
&& A_{\pm} \;\Rightarrow \; \tilde{A}_\pm = \left(1 \pm
a\,\beta_0\right)\left(1 -2 \alpha_0\,\lambda a c \,h\right)
\pm 2 \lambda a^2 c\,h~, \nn
&& B_{\pm} \; \Rightarrow \;
\tilde{B}_\pm =  \left[1 \pm \frac{\alpha_0}{a}\left(1 \mp a
\beta_0\right)\right] B_{\pm} - a\left[\beta_0 +
\frac{\alpha_0}{a^2}\left(1 \mp a\beta_0 \right)\right] h~.
\eeqa
The appearance of an additional phase factor in \p{sol2} is due to the fact
that in the general case the $\varepsilon$ transformations \p{tranFVKP} acquire
the common extra piece proportional to $\beta_0$:
$$
\delta_\varepsilon {\cal F} \;\Rightarrow \; \delta_\varepsilon {\cal F}-
i\varepsilon \,\beta_0 a {\cal F}~,
$$
etc. The QK Taub-NUT and QK EH truncations of the general solution correspond
to imposing the following conditions:
\beqa
&& \underline{\mbox{QK Taub-NUT}}: \qquad \beta_0 = 0~, \quad c=0~, \quad,
\alpha_0 c \equiv \tilde{\alf}_0  \neq 0~, \label{TNcond} \\
&& \underline{\mbox{QK EH}}: \qquad \qquad \;\; \;\;\;\alpha_0 = 0~, \quad
\beta_0\,a \equiv \tilde{\be}_0 \neq 0~, \quad a\; \Rightarrow \; 0~.
\label{EHcond} \eeqa
Respectively, in these two limits we have
\beqa
&&\underline{\mbox{QK Taub-NUT}}: \nn
&&\tilde{A}_{\pm} = (1-2\lambda a \tilde{\alf}_0 h)
\equiv \tilde{A}~, \quad \tilde{B}_{\pm} = \pm \left[ h +
{\tilde{\alf}_0\over a}(1 + \lambda a^2 r^2)\right] \equiv \pm \tilde{B}~,
\nn && \tilde{\Delta}_{+} = \tilde{\Delta}_{-} = \tilde{B}^2 +
t^2\tilde{A}^2 \; \Rightarrow \; \tilde{L} = \tilde{M}~, \quad \tilde{R} =
\tilde{S}~, \label{TNsol} \\
&&{} \nn
&&\underline{\mbox{QK EH}}: \nn
&& \tilde{A}_\pm = (1\pm \tilde{\be}_0)~, \quad \tilde{B}_\pm = c \pm
(1\mp\tilde{\be_0})h~, \nn
&& \tilde{\Delta}_{\pm} = \left[ c \pm (1\mp \tilde{\be_0})\right]^2 + (1\mp
\tilde{\be}_0)^2 t^2~. \;  \label{EHsol}
\eeqa
Note that in the Taub-NUT case we can obviously choose, up to a gauge
freedom,
$$
{\cal P} = {\cal V}~, \quad {\cal F} = {\cal K} \;\Rightarrow \; e^{2i\mu} =
-ie^{i\alpha}~,
$$
which, according to the above definition of ${\cal P},\,{\cal V},\,{\cal
F},\,{\cal V}$, corresponds just to the truncation  $Q^{+a}_{A=2} = 0$
at the level of the general HSS Lagrangian \p{actionGen}. Also note that
for taking the QK EH limit in the original form  of constraints
\p{constr11a}, \p{constr22a} in the unambiguous way, one should firstly
rescale $g^{ik} \rightarrow a \,g^{(ik)}~.$ The corresponding limiting QK
metrics can be obtained by taking the limits \p{TNcond}, \p{EHcond} in the
QK metric associated with the general choice of $\alpha_0 \neq 0~, \;\beta_0
\neq 0~.$

\setcounter{equation}{0}
\section{The structure of general metric}
\subsection{First set of coordinates}
To obtain the metric, we substitute the explicit form \p{sol3} of the
coordinates into the distance \p{dist00} and compute it. The algebraic
manipulations to be done in order to cast the resulting expression in a
readable form are rather involved, and Mathematica was intensively used while
doing this job. To simplify matters, we make the change of coordinates
\beq\label{chg1}
T=\frac{2\,t}{1- a^2\la\,r^2}~,\qq H=\frac{2\,h}{1-a^2\la\,r^2}~,\qq
\rho=\sqrt{T^2+H^2}\,
\eeq
\noindent and use the notations
$$\beta=\frac{a\,\beta_0}{1 - 4\,c\,\lambda }~,\qq
  c_\pm = \left(\frac{1}{1 \mp a\,\beta_0} \pm \frac{\alpha_0}{a}
\right)\,c~,$$
$$\delta_\pm=\frac{4\,\Delta_\pm}{\left(1-a^2\la\,r^2\right)^2}=
\left(1+4\,a^2\la\,c^2_\pm\right)T^2+\left(H \pm 2\,c_\pm\right)^2\, . $$
The final result for the metric $\EuFrak{g}$ can be presented in terms
of 4 functions $D,\; A,\;P,\; Q$
\beq\label{met1}
4 D^2\,\EuFrak{g}=\frac{P}{A} \left(d\phi+\frac{Q}{4\ P}d\alf\right)^2+ A
\left(\EuFrak{g}_0+\frac{1+a^2\la\,\rho^2}{P}\,T^2\,d\alf^2\right)
\eeq
\noindent where
\beq\label{h2}
\EuFrak{g}_0=\frac{dH^2+dT^2+a^2\la\,(T\,dH-H\,dT)^2}{1+a^2\la\,\rho^2}\eeq
is the metric on the two-sphere ($a^2\la<0$), on flat space ($a^2\la=0$), or on
the
hyperbolic plane ($a^2\la>0$).

The various involved functions are as follows
\beqa
D &=&\frac{\cal D}{1-a^2\la\,r^2}\ =\  1 - \lambda \,\left( \left( 1 +
a\,\beta_0 \right) \,{\sqrt{{{\delta }_-}}} +
     \left( 1 - a\,\beta_0 \right) \,{\sqrt{{{\delta }_+}}} \right)\ \ ,\nn[5mm]
A &=&\frac{a^2}{4} + \frac{1}{4}\,\left (\left( 1 + a\,\beta_0 \right) \,\frac{
        1 - 4\,a^2\,\lambda \,{{c_-}}^2  }{{\sqrt{{{\delta }_-}}}} +
     \left( 1 - a\,\beta_0 \right) \,\frac{1 - 4\,a^2\,\lambda \,{{c_+}}^2  }
      {{\sqrt{{{\delta }_+}}}}\right)\nn
&&+\,a^2\,\lambda\,H \,\left( \frac{\left( 1 + a\,\beta_0 \right)
\,{c_-}}{{\sqrt{{{\delta }_-}}}} -
     \frac{\left( 1 - a\,\beta_0 \right) \,{c_+}}{{\sqrt{{{\delta }_+}}}}
\right)  -
  \frac{4\,c^2\,\lambda}{1 - a^2\,{\beta_0}^2}
   \frac {1 + a^2\,\lambda\,H^2}{{\sqrt{{{\delta }_-}}}\,{\sqrt{{{\delta
}_+}}}}\ \ ,\nn[5mm]
P &=& \left( 1 + a^2\,\lambda \,{\rho }^2 \right) \,
  {\left( 1 - 2\,c\,\lambda \,\frac{H + 2\,{c_+}}{{\sqrt{{{\delta }_+}}}}
    +2\,c\,\lambda \,\frac{H - 2\,{c_-}}{{\sqrt{{{\delta }_-}}}}  \right) }^2\nn
&&+\,4\,c^2\,{\lambda }^2\,T^2\,{\left( -a\,\alpha_0  -
      \frac{1 + 2\,a^2\,\lambda \,{c_-}\,H}{{\sqrt{{{\delta }_-}}}} +
      \frac{1 - 2\,a^2\,\lambda \,{c_+}\,H}{{\sqrt{{{\delta }_+}}}} \right) }^2\
\ ,\nn[5mm]
Q &=& -\left( 1 + a^2\,\lambda \,{\rho }^2 \right) \,
  \left( 2\,\beta \,\left( 1 + 2\,c\,\lambda  \right)  +
    \left( 1 + \beta \,\left( 1 + 4\,c\,\lambda  \right)  \right) \,\frac{ H -
2\,{c_-}}
     {{\sqrt{{{\delta }_-}}}} \right.\nn
&& \left. +\,\left( 1 - \beta \,\left( 1 + 4\,c\,\lambda  \right)  \right)
\,\frac{H + 2\,{c_+}}
   {{\sqrt{{{\delta }_+}}}} - 4\,c\,\beta \,\lambda \,\frac{\left( H - 2\,{c_-}
\right) \,
     \left( H + 2\,{c_+} \right) }{{\sqrt{{{\delta }_-}}}\,{\sqrt{{{\delta
}_+}}}}\right)\nn
&&-\,2\,c\,a\,\lambda \,T^2\,\Bigg( a^2\,\alpha_0 -
    2\,\left( 1 - 2\,c\,{\alpha_0}^2 \right) \,\beta \,a\,\lambda\nn
&&\hspace{2.3cm}  + \left( a + \alpha_0 + \alpha_0\,\beta \,\left( 1 +
4\,c\,\lambda  \right)
         \right) \,\frac{ 1 + 2\,{c_-}\,a^2\,\lambda\,H}{{\sqrt{{{\delta }_-}}}}
\nn
&&\hspace{2.3cm}  -\, \left( a - \alpha_0 + \alpha_0\,\beta \,\left( 1 +
4\,c\,\lambda  \right)
       \right) \,\frac{1 - 2\,{c_+}\,a^2\,\lambda\,H}{{\sqrt{{{\delta }_+}}}}\nn
&&\hspace{2.3cm}\left. - \frac{2\,\beta}{a}
  \frac{\left( 1 + 2\,{c_-}\,a^2\,\lambda\,H \right) \,\left( 1 -
2\,{c_+}\,a^2\,\lambda\,H \right) }
   {{\sqrt{{{\delta }_-}}}\,{\sqrt{{{\delta }_+}}}}\right)\ .
\label{functions}
\eeqa
The isometry group $U(1)\times U(1)$ acts by translations on $\phi$ and $\alf$.

\subsection{Second set of coordinates}
In order to verify that $\EuFrak{g}$ is self-dual Einstein (see section 5.4),
it is more convenient to use coordinates $s$ and $x$ defined by
\beq\label{chg2}
T=s\,\sqrt{1-x^2}~,\qq H=s\,x\, .
\eeq
We then get for the metric the expression
\beq\label{met2}
\barr{l}
\dst 4 D^2\,\EuFrak{g}=\frac{P}{A} \left(d\phi+\frac{
Q}{4\ P}d\alf\right)^2\\[5mm]\dst
\hspace{3cm}+A\,\left(\frac{ds^2}{1+a^2\la\,s^2}+\frac{s^2\,dx^2}{1-x^2}
+\frac{s^2\,\left(1+a^2\la\,s^2\right)\left(1-x^2\right)}{P}\,d\alf^2\right)~.\earr
\eeq
The functions $A$, $P$, $Q$ and $D$ are still the same as in
(\ref{functions}), up to the substitution (\ref{chg2}), and the functions
$\de_{\pm}$ can be written as
$$\de_{\pm}=\frac 1{a^2\la}\left[(1+4a^2\la c_{\pm}^2)(1+a^2\la s^2)
-(1\mp 2a^2\la c_{\pm}sx)^2\right].$$

\subsection{Third set of coordinates ($\alpha_0 = \beta_0= 0$ case)}
In the limit $\alpha_0 \rightarrow 0$ and $\beta_0 \rightarrow 0$, the metric
$\EuFrak{g}$ reduces to the quaternionic extension of the double Taub-NUT
metric given in \cite{civ}. For this particular case one can
get rid of the square roots by switching to the spheroidal
coordinates ($u,\theta$),
\beq\label{chg3}
T=\frac{\sqrt{u^2-4\,c^2}}{\sqrt{1+4\,a^2\la\,c^2}}\,\sin\theta~,\qq
H=u\,\cos\theta~.
\eeq
In these coordinates:
$$\sqrt{\delta_\pm}=u \pm 2\,c\,\cos\theta \ .$$

It is convenient to scale the angles $\phi$ and $\alf$ according to
$$\widehat \phi=\frac{\phi}{1+4\,a^2\la\,c^2}~,\qq\widehat
\alf=\frac{\alf}{1+4\,a^2\la\,c^2}\ .$$
Then the metric at $\alpha_0 = \beta_0 =0$ becomes
\beq
\begin{array}{rrl}
\dst 4 D^2\,\EuFrak{g} &=& \dst (1+a^2\la\,u^2)\,\frac{\widehat P}{\widehat A}
\left(d\widehat\phi+\frac{\widehat Q}{4\widehat P}d\widehat\alf\right)^2\\[5mm]
&&\hspace{2.5cm}\dst + \widehat A\,
\left({\widehat{\EuFrak{g}}}_0+\frac{(u^2-4c^2)(1+4\,a^2\la\,c^2\cos^2\theta)}{\widehat P}
\sin^2\theta\,d\widehat\alf^2\right),
\end{array}
\label{met3}
\eeq
where
$${\widehat{\EuFrak{g}}}_0=(u^2-4\,c^2\cos^2\theta)\,\EuFrak{g}_0=\frac{du^2}{(u^2-4c^2)(1+a^2\la\,u^2)}+\frac{d\theta^2}{1+4\,a^2\la\,c^2\cos^2\theta}$$
\noindent and
\beqa
4\,\widehat A &=& 4\,(u^2-4\,c^2\cos^2\theta)\,A\nn
&=& (2+a^2u)(u-8\,c^2\la)-4\,a^2c^2D^2\cos^2\theta~,\nn[5mm]
\widehat P &=&
\frac{(1+4\,a^2\la\,c^2)(u^2-4\,c^2\cos^2\theta)}{1+a^2\la\,u^2}\,P\nn
&=& 4\,c^2\sin^2\theta\,(1+4\,a^2\la\,c^2\cos^2\theta)\,
D^2+(u^2-4\,c^2)(1+4\,a^2\la\,c^2\cos^2\theta
-16\,\la^2c^2\sin^2\theta)~,\nn[5mm]
\widehat Q &=&
\frac{(1+4\,a^2\la\,c^2)(u^2-4\,c^2\cos^2\theta)}{1+a^2\la\,u^2}\,Q\nn
&=& -2(u^2-4\,c^2)(1+4\,a^2\la\,c^2)\,\cos\theta~,\nn[5mm]
D&=&1-2\,\la\,u~.
\label{functions3}
\eeqa

\subsection{Einstein and self-dual Weyl properties of the metric}
A four-dimensional QK metric is nothing but an Einstein metric
with self-dual Weyl tensor.
This property should be inherent to the metric $\EuFrak{g}$ given by
(\ref{met1}),
since we started from the generic HSS action for QK sigma models. However,
checking
these properties explicitly is a good test of the correctness of our
computations.

We first consider the particular case $\,\alf_0=\be_0=0\,$  because
the use of  the spheroidal-like  coordinates (\ref{chg3}) greatly
simplifies the metric as can be seen from relations (\ref{met3}) and
(\ref{functions3}). Despite these simplifications, intensive use of
Mathematica was needed to compute the spin connection, the
anti-self-dual curvature $\,R^-_i\,$ and to check the crucial relation
(see the Appendix for the notation) :
$$R^-_i=-16\,\la\ \Xi^-_i~.$$
It simultaneously establishes that the metric is indeed
self-dual Einstein, with $$ Ric \,(g) =\Lambda\ g~,\qq\quad
\frac{\Lambda}{3}=-16\la~,\qq\qq W^-_i=0~.$$

For non-vanishing $\,\alf_0\,$ or $\,\be_0\,$, such a check is no
longer feasible because of the strong increase in complexity
of various functions appearing in the metric. Moreover, in this
case we failed to find any proper generalization of the spheroidal-like
coordinates (\ref{chg3}) which would allow us to get rid of the square roots
$\,\sqrt{\de_+}\,$ and $\,\sqrt{\de_-}~.$

In order to by-pass these difficulties we have used an approach due to
Przanowski \cite{Pr} and Tod  \cite{To}, which reduces the verification
of the self-dual Einstein property to simpler checks.
We shall begin with a description of their construction.

One starts from
an Einstein metric $g$ (more precisely, $\,Ric\,(g)=\Lambda\,g).$ Furthermore
it will be supposed that this metric has (at least) one Killing
 vector with the
associated 1-form $\,K=K_{\mu}\,dx^{\mu}~.$ Differentiating $K$ gives
$$dK=dK_i^+\,\Xi_i^++dK_i^-\,\Xi_i^-\,,\qq\Xi_i^\pm=e_0\wedge e_i \pm \frac12
\eps_{ijk}\,e_j\wedge e_k\ ,$$
for some vierbein of the metric $g.$
We can extract, from $dK$,  an integrable complex structure ${\cal I}$ and a
coordinate $w$ according to
\beq\label{sc}
{\cal I}=\frac{dK^-_i}{\dst\sqrt{\sum_i {(dK_i^-)^2}}}\ \Xi^-_i~,\qq
w=-\frac{\Lambda}{3\sqrt{\dst\sum_i {(dK_i^-)^2}}}~.\eeq
Using these elements one can formulate
\begin{nth}[\cite{Pr},\cite{To}]
There exist real coordinates $\,w,\,\nu\,$ and $\,\mu\,$ such that any Einstein
metric $\,g\,$ with self-dual Weyl tensor and a Killing vector $\,\pt_{\phi}\,$
can be written as
\beq\label{PT}
g=\frac 1{w^2}\left[\frac 1{\cal W}(d\phi+\Theta)^2+
{\cal W}(e^v(d\nu^2+d\mu^2)+dw^2)\right].\eeq
This metric will be self-dual Einstein iff
\beq\label{tod}
\left\{\barr{crcl}
\dst ({\rm a}) &\dst\  -2\,\frac{\Lambda}{3}\,{\cal W} &=&
2-w\,\partial_w\,v~,\\[5mm] \dst ({\rm b}) &\dst\ \left(\partial^2_\nu
+\partial^2_\mu\right)\,v+ \partial^2_w \,(e^v)&=&0~,\\[5mm]
\dst ({\rm c}) &\dst -d\Theta =\partial_\nu\,{\cal W}\,d\mu\wedge dw &+&
\partial_\mu\,{\cal W}\,dw\wedge d\nu+
\partial_w\,\left({\cal W}\,e^v\right)\,d\nu\wedge d\mu~. \earr\right.
\eeq\end{nth}

\noindent The following remarks are in order :
\brm
\item The relation (\ref{tod}b) is the celebrated continuous Toda equation.
\item Except for this Toda equation, the checks of the self-dual Einstein
property are reduced to solving first order partial differential equations.
\item Relation (\ref{PT}) shows that any self-dual Einstein metric with at
least one Killing is conformal to a subclass of K\" ahler scalar-flat metrics
(see section 6.1 for the proof).
\erm

Let us now use this approach to analyze our metric (\ref{met2}) in the
($s,\,x$) coordinates and to check whether it obeys the conditions \p{tod}.

We take for vierbein $$\begin{array}{ll}
\dst e_0 = \frac1{\sqrt{W}}\,\left(d\phi+\Theta\right)~, \qq&
\dst e_1 = \frac{\sqrt{A}}{2\,D}\,\frac{ds}{\sqrt{1+a^2\la\,s^2}}~, \\[5mm]
\dst e_2 = \frac{\sqrt{A}}{2\,D}\,\frac{s\,dx}{\sqrt{1-x^2}}~,\qq &
\dst e_3 =
\frac{\sqrt{W}}{4\,D^2}\,s\,\sqrt{1+a^2\,\la\,s^2}\,\sqrt{1-x^2}\,d\alf\ ,
\end{array}\qq W=\frac{4D^2A}{P}\ ,
$$
and consider the Killing $\partial_\phi~,$ with the 1-form
$$K=\frac1{W}\,\left(d\phi+\Theta\right)=\frac1{\sqrt{W}}\,e_0\ .$$
The computation of $\dst \sum_i (dK_i^-)^2$ eventually leads to the
identification \beq\label{w}
w=-\frac{\Lambda}{3}\,\frac{D}{4\,\la\,\sqrt{\delta({\hat c})}}~,
\eeq
where
\beq
\label{delta}
\begin{array}{rll}
\dst \delta({\hat c}) &=&\dst \frac{1}{a^2\,\lambda}\left[{
   \left( 1 +  4\,a^2\,\lambda \,{\hat c }^2 \right)\,\left( 1 + a^2\,\lambda
\,s^2 \right)-\left( 1 -2\, a^2\,\lambda\,\hat c \,s\,x  \right)^2
}\right]\,,\\[2mm]
\dst 2\,\hat c &=&\dst \left( 1 - a\,\beta_0 \right)\,{c_+}
-\left( 1 + a\,\beta_0 \right) \,{c_-}
          =2\,c\frac{\alf_0}{a}\,.
\end{array}
\eeq
Then, comparing the metric $\EuFrak{g}$ in the form (\ref{met2})
with (\ref{PT}), we express the quantities ${\cal W},\,\mu\,$
and $e^v$ entering \p{PT} in terms of ours
\beq\label{id1}{\cal W}=\frac W{w^2}~,\qq\mu=\alf\ ,\qq e^v=
\frac{s^2\,\left( 1 + a^2\la\,s^2 \right) \,
\left( 1 - x^2 \right)\,w^4 }{16\,D^4}\ .\eeq
Simultaneously, we obtain the expressions for the partial derivatives of $\nu$
\beq\label{id2}
\partial_x\,\nu=-\frac{4\,D^2\,\partial_s\,w}{\left( 1 - x^2 \right)\,w^2 }\
,\qq \partial_s\,\nu=\frac{4\,D^2\,\partial_x\,w}{s^2\,\left( 1 + a^2\la\,s^2
\right) \,w^2}\ .\eeq
Two expressions for the mixed derivative $\partial_s\partial_x\nu$
coincide  as a consequence of the relation:
\beq\label{id3}
s^2(1 + a^2\la\,s^2)\partial^2_s\,D+(1-x^2)\partial^2_x\,D=0\ .\eeq

Checking the relation (\ref{tod}a) suggests the identification
\beq\label{id4}\frac{\Lambda}{3}=-16\,\la\ .\eeq
Then the remaining equations (b) and (c) in (\ref{tod}) have been
explicitly checked  using Mathematica, and shown to be valid.
This proves that our general metric \p{met2} is self-dual
Einstein.

\subsection{Limiting cases}
\noindent\underline{\it The hyper-K\" ahler limit}\\

Using the coordinates $H$ and $T$ (defined in \ref{chg1}), in the limit
$\la\rightarrow 0$, the metric (\ref{met1}) can be written as the multicentre
structure $$4\,\EuFrak{g}(\la\rightarrow
0)=\frac1{V}\,\left(d\Phi+\cal{A}\right)^2+V\,\EuFrak{g}_0(\la\rightarrow 0)\
,$$
\noindent with the flat 3-metric and the angle $\Phi$ defined by
$$\EuFrak{g}_0(\la\rightarrow 0)=dH^2+dT^2+T^2d\alf^2\
,\qq\Phi=\phi-\frac{a\,\beta_0}2\,\alf\ .$$
\noindent The potential $V$ and the connection $\cal A$ are, respectively,
\beq\label{pot1}
V=\frac14 \,
\left(a^2+\frac{1+a\,\beta_0}{\sqrt{\delta_-}}+\frac{1-a\,\beta_0}{\sqrt{\delta_+}}\right)\,,\eeq
\beq\label{pot2}
{\cal A}=-\frac14
\left(\left(1+a\,\beta_0\right)\,\frac{H-2\,c_-}{\sqrt{\delta_-}}+\left(1-a\,\beta_0\right)\,\frac{H+2\,c_+}{\sqrt{\delta_+}}\right)\,d\alf\,,\eeq
\noindent with
$$\delta_\pm=\ \left(H \pm 2c_\pm\right)^2+T^2\ ,\qq c_\pm =\frac{c}{1 \mp
a\,\beta_0}\ .$$
Since $\alf_0$ is an irrelevant parameter in the limit $\la\rightarrow 0$ (it
can be removed from the metric by a shift of $H$), we put it equal to zero
from the very beginning.

The potential shows two centres at $\dst T=0~,\, H=\mp2\,c_\pm $ with different
masses $\dst\frac{1\mp a\,\beta_0}4$ and $\dst V(\nf)=\frac{a^2}4$ . An easy
computation gives the fundamental multicentre relation
$$dV=-\,\mathop{\smash{\star}}_{\EuFrak{g}_0(\la\rightarrow0)}\,d{\cal A}\ .$$
\noindent  For $\dst a\neq0,\,\beta_0=0$, we have the double Taub-NUT metric;
for $\dst a\neq0,\,c=0$ and $\dst a\neq0,\,a\,\beta_0=\pm 1,$ ( $c_\pm $
finite), we have the Taub-NUT metric; for $\dst a=0$, we have the
Eguchi-Hanson metric.
\vspace{0.4cm}

\noindent\underline{\it The quaternionic Taub-NUT limit}\\

In order to show that in the limit $c\rightarrow0$ we recover the quaternionic
Taub-NUT metric, we switch to new  coordinates ($\hat s$, $\hat \theta$)
defined by  $$H=\frac2a\frac{\hat s\,\cos\hat \theta}{1-\la\,\hat s^2}\ ,\qq
T=\frac2a\frac{\hat s\,\sin\hat \theta}{1-\la\,\hat s^2}\ .$$
The metric $\EuFrak{g}$ coincides, up to a constant factor
$\dst\frac1{2\,a}$, with the metric given by relation (5.4) in \cite{iv1} :
$$2\,a\,\EuFrak{g}(c\rightarrow 0)=\frac1{2}\left(\frac{\hat B}{\hat s\,{\hat
C}^2}\,d\hat s^2+\frac{\hat s\,\hat B}{{\hat C}^2}\,(\si_1^2+\si_2^2)+\frac{\hat
s {\hat A}^2}{\hat B\,{\hat C}^2}\,\si_3^2\right)\ ,$$
\noindent where
$$\hat A=1-\hat R\,{\hat \la}^2{\hat s}^2\ ,\qq\hat B=1+{\hat \la}^2\hat
s\,(4+\hat R\,\hat s)\ ,\qq\hat C=1+\hat R\,\hat s+\hat R\,{\hat \la}^2{\hat
s}^2\ ,$$
\noindent and
\beq
\left\{
\begin{array}{l}
\dst\si_1^2+\si_2^2 =d{\hat\theta}^2+\sin^2\hat\theta\,d\alf^2\ ,\\[2mm]
\si_3=(-2\,d\phi+a\,\beta_0\,d\alf)+\cos\hat\theta\,d\alf\ ,\\[2mm]
\dst\hat R=-4\,\frac{\la}{a}\ ,\qq{\hat\la}^2=\frac a4\ .
\end{array}
\right.\eeq
\noindent Various limits of the quaternionic Taub-NUT metric can be found in
\cite{iv1}. Let us just remark here that in the limit $\hat R\rightarrow0$
we once again recover the standard Taub-NUT metric.
\vspace{0.4cm}

\noindent\underline{\it The quaternionic Eguchi-Hanson limit}\\

In the limit $a\rightarrow0$ with $a\,\beta_0=\tb\neq0$, it is more convenient
to study the metric in coordinates in which the square roots disappear.
Thus, we define the coordinates $\tilde s$ and $\tilde \theta$ by
$$T=\frac{2}{1-\tb^2}\,\sqrt{\ts^2-c^2}\sin\tth\ ,\qq
H=\frac{2}{1-\tb^2}\,\ts\,\cos\tth+c_--c_+\ ,$$
so that
$$\sqrt{\delta_\pm}=\frac{2}{1-\tb^2}\,(\ts\pm c\,\cos\tth)\ .$$
The metric can now be expressed as
$$4\,(1-\tb^2)\,\tilde
C^2\,\EuFrak{g}(a\rightarrow0,\,\tb)=\frac{\ts^2-c^2}{\ts\,\tilde B}\,{\cal
G}^2+\ts\,\tilde B\,\left(\frac{d\ts^2}{\ts^2-c^2}+d\tth^2+\sin^2\tth\,{\cal
H}^2\right)\ ,$$
with
$$\begin{array}{rcll}
\dst\tilde C &=& 1-\frac{\kappa^2}{1-\tb^2}\,\left(\ts-c\,\tb\,\cos\tth\right),
& \dst\ {\cal G}
= -\left(1+\beta\,\cos\tth\right)d\alf+2\,\cos\tth\,d\phi\ ,\\[5mm]
\dst\ts\,\tilde B &=& \ts-\kappa^2\,c^2+c\,\tb\,\cos\tth~,&
\ {\cal H} =\dst\frac{1}{\ts \tilde
B}\,\left[-\left(\ts-c\right)\,\beta\,d\alf+2\left(\ts-\kappa^2c^2\right)\,d\phi\right]\,\end{array}
$$
where
$$\kappa^2=4\,\la\ ,\qq\beta=\frac{\tb}{1-4\,c\,\la}\ .$$
One can see that in the limit $a\rightarrow0$, the parameter
$\alf_0$ fully drops out from the metric. If we now take the limit
$\tb=a\,\beta_0\rightarrow0$, we reproduce the quaternionic
Eguchi-Hanson metric derived in \cite{IvV}  (see equation (4.7) of this
reference) : $$4\,\tilde
C^2\,\EuFrak{g}(a\rightarrow0)=\frac{\ts^2-c^2}{\ts\,\tilde
B}\,{\tilde\si_3}^2+\ts\,\tilde B\,\left(\frac{d\ts^2}{\ts^2-c^2}+{\tilde
\si_1}^2+{\tilde \si_2}^2\right)\ ,$$ with
$$\begin{array}{rllrll}
\tilde C &=&\dst 1-\kappa^2\,\ts\ ,&
\tilde\si_3 &=& \left(-d\alf\right)+\cos\tth\,\left(2\,d\phi\right)\ ,\\[5mm]
\dst\ts\,\tilde B &=&\dst \ts-\kappa^2\,c^2\ ,&
{\tilde \si_1}^2+{\tilde \si_2}^2&=&\dst
d\tth+\sin^2\tth\,\left(2\,d\phi\right)^2\ .
\end{array}$$
In conclusion, let us point out that, whereas the parameters $a,\ c\ $ and
$\be_0$ have a counterpart in the HK limit, this is not the case
for the parameter $\alf_0.$ This distinguished parameter is specific just
for the QK metrics.

\setcounter{equation}{0}
\section{Connection with the literature}
Metrics with self-dual Weyl tensor may appear as :
\brm
\item K\" ahler scalar-flat metrics,
\item Self-dual Einstein metrics (considered in this work),
\item Metrics in the system of coupled Einstein-Maxwell fields.
\erm
In order to exhibit the relationships between these classes and to
find out how our metrics correlate with them, let us begin with
the description, due to LeBrun, of the K\" ahler scalar-flat metrics with one
Killing vector.

\subsection{K\" ahler scalar-flat metrics in LeBrun setting}
These metrics, with self-dual Weyl tensor, have received attention in
\cite{Le}. There, it was proved that any such metric,
with at least one Killing vector $\,K=\partial_t~,$ can be written as
\beq\label{lb1}
g=\frac 1{\cal W}(dt+\tilde\Theta)^2+{\cal
W}[dw^2+e^v(d\nu^2+d\mu^2)]=\sum_{A=0}^3\,e_A^2~,\eeq
where the functions $\,v\,$ and $\,{\cal W}\,$ must be solutions of the
following equations  \beq\label{lb2}
(\partial_{\nu}^2+\partial_{\mu}^2)v+\partial_w^2(e^v)=0~,\qq
(\partial_{\nu}^2+\partial_{\mu}^2){\cal W}+\partial_w^2({\cal
W}\,e^v)=0~.\eeq
The connection one-form $\,\tilde\Theta\,$ is then obtained from
\beq\label{lb3}
d\tilde\Theta=\partial_{\nu}({\cal W})\,d\mu\wedge dw+\partial_{\mu}({\cal
W})\,dw\wedge d\nu+
\partial_{w}({\cal W}e^v)\,d\nu\wedge d\mu~.\eeq
The vierbein, defined in relation (\ref{lb1}), is taken to be
$$ e_0=\frac{dt+\tilde\Theta}{\sqrt{{\cal W}}}~,
\qq e_1=\sqrt{{\cal W}e^v}\,d\nu~,\qq e_2=\sqrt{{\cal W}e^v}\,d\mu~,\qq
e_3=\sqrt{{\cal W}}\,dw~.$$
In terms of the self-dual two-forms
$\dst\ \Xi^{\pm}_i=e_0\wedge e_i\pm\frac 12\,\eps_{ijk}\,e_j\wedge e_k\ $
the K\" ahler form is anti-self-dual,
\beq\label{omega}
\Omega=dw\wedge(dt+\tilde\Theta)+{\cal W}e^v\,d\nu\wedge d\mu=-\Xi_3^-~,
\eeq
while the Ricci form is self-dual,
\beq\label{rho}
2{\hat \rho}=\frac 1{\sqrt{e^v}}\,\partial_{\nu}\left(\frac{\partial_{w}v}{{\cal
W}}\right)\cdot\Xi_1^+
+\frac 1{\sqrt{e^v}}\,\partial_{\mu}\left(\frac{\partial_{w}v}{{\cal
W}}\right)\cdot\Xi_2^+
+\partial_w\left(\frac{\partial_{w}v}{{\cal W}}\right)\cdot\Xi_3^+~.
\eeq
Now, comparing (\ref{PT}) and (\ref{lb1}), we observe that any self-dual
Einstein metric with at least one Killing, in particular the metric \p{met1},
is conformally related to a subclass of K\" ahler  scalar-flat metrics, with
the identifications : $$
\tilde\Theta=-\Theta~,\qq \qq dt=-d\phi~,\qq \qq d\mu=d\alpha~,\qq \qq
g=w^2\,\EuFrak {g}~.
$$

In \cite{Le}, a large class of explicit solutions of (\ref{lb2}) was obtained.
Taking
$$q=\sqrt{2w}~,\qq e^v=q^2~,\qq
V=q^2W~,\qq\qq\ga=\frac{d\nu^2+d\mu^2+dq^2}{q^2}~,\qq $$
where $\,\ga\,$ is the hyperbolic 3-space, these metrics have the form
\beq
q^2\left[\frac 1V(dt+\Theta)^2+V\,\ga\right], \label{part}
\eeq
where $\,V\,$ is some real harmonic function on $\,\ga.$

LeBrun obtained the potential $\,V\,$ as a sum of monopoles in this
hyperbolic space. In the limit where the hyperbolic space becomes flat, one
recovers the multicentre metrics. However, the possibility that these
metrics could be conformally Einstein has been ruled out by Pedersen
and Tod in \cite{pt}. Therefore the metrics \p{part} bear no relation to our
metric (\ref{met1}).

\subsection{Flaherty's equivalence}

Let us now examine Flaherty's equivalence relating K\" ahler
scalar-flat metrics and self-dual metrics solving the coupled
Einstein-Maxwell field equations.

In \cite {Fl} Flaherty has proved :

\begin{nth}
The following two classes of metrics are equivalent :
\brm
\item Any K\" ahler scalar-flat metric.
\item Any metric which is a solution of the coupled  Einstein-Maxwell equations
\beq\label{EM}
\left\{\barr{l}
Ric_{\,\mu\nu}=\frac 12\left(F_{\mu\rho}g^{\rho\si}F_{\nu\si}
-\frac 14\,g_{\mu\nu}\,F_{\rho\si}\,F^{\rho\si}\right),\\[5mm]
dF^-=0~,\qq dF^+=0~,\earr\right.
\eeq
with self-dual Weyl tensor ($\,W^-=0\,$).\erm

\noindent In this equivalence the self-dual parts of the Maxwell field
strength are given by
$$F^-\propto\Omega~,\qq\quad F^+\propto{\hat \rho}~,$$
where $\,\Omega\,$ denotes the K\" ahler form and $\,{\hat \rho}\,$ the Ricci
form of
the K\" ahler metric.
\end{nth}
In the euclidean case, this equivalence can be easily checked for metrics
with at least one Killing vector, using the LeBrun
framework. One can check eqs. (\ref{EM}) and find the self-dual parts of
the field strength two-forms :
\beq\label{max}
 \dst F^-=-\frac{m}{2}\ \Omega~, \qq\qq F^+=\frac 2m\ {\hat \rho}~,
\eeq
where $\,m$ is an arbitrary real parameter.

This equivalence and the property that any self-dual Einstein
metric with one Killing is conformal to some K\"ahler scalar-flat metric
suggest that the Weyl-self-dual metrics which solve the
Einstein-Maxwell system may hide, up to some conformal factor, a  self-dual
Einstein  metric. Let us now examine two known classes of the metrics giving
solution of the Einstein-Maxwell system (in general, they are not
Weyl-self-dual) in order to see whether the metric (\ref{met1}) is
conformally related to any of them. We shall find that the answer is negative
in both cases. This means that \p{met1} determines a new explicit solution of
the Einstein-Maxwell system, with the conformal factor $w$ given in (\ref{w}).

\subsection{The metrics of Perj\`es-Israel-Wilson}

These metrics are solutions of the Einstein-Maxwell field equations.
They were derived independently, for the minkowskian signature, by Perj\`es
\cite{Pe} and Israel and Wilson \cite{iw}. Their continuation to the
euclidean signature  was given by Yuille \cite{Yu} and Whitt \cite{Wh} who
discussed  their global properties and their possible applications in the
path integral approach to quantum gravity.

These metrics have at least one Killing vector $\,\partial_t.$
Their local form is given by
\beq\label{ipw1}
g=\frac 1V(dt+{\cal A})^2+V\,\ga_0~,\,\qq V=U\,\tilde{U}~,
\qq \ga_0=d\vec{x}\cdot d\vec{x}~.\eeq
The real functions  $\,U\,$ and $\,\tilde{U}\,$ must be harmonic
\beq\label{ipw2}
\Delta\,U=\Delta\,\tilde{U}=0~,\eeq
and the connection one-form $\,{\cal A}\,$ is constrained by
\beq\label{ipw3}
\stdec_{\ga_0}\,d{\cal A}=\tilde{U}\,dU-U\,d\tilde{U}~.\eeq
The star and laplacian are taken with respect to the three dimensional
flat space with cartesian coordinates $\,\vec{x}~.$ Clearly, when
$\,U\,$ or $\,\tilde{U}\,$ are constant we come back to the multicentre
metrics.

In order to check the previous assertions, let us define the vierbein
$\,e_A\,$ by
$$e_0=\frac 1{\sqrt{V}}\,(dt+{\cal A})~,\qq\qq e_i=\sqrt{V}\,dx_i~,
\qq\qq i=1,2,3~.$$

It is an easy task to compute the matrices $\,A,\,B\,$ and $\,C\,$ giving the
curvature (see the Appendix for the definitions and notation). One finds,
upon using the relations (\ref{ipw2}), (\ref{ipw3}), the simple expressions
\beq\label{weyl1}
\barr{l}
\dst A_{ij}=\frac1V\left[\frac{\pt^2_{ij}U}{U}
-3\frac{\pt_i U\pt_j U}{U^2}+\delta_{ij}\frac{(\pt_l U)^2}{U^2}\right],\\[5mm]
\dst B_{ij}=-\frac 1{V^2}\,\pt_i U\,\pt_j \tilde{U}~,\\[5mm]
\dst C_{ij}=\frac 1V\left[\frac{\pt^2_{ij}\tilde{U}}{\tilde{U}}
-3\frac{\pt_i\tilde{U}\pt_j\tilde{U}}{\tilde{U}^2}
+\delta_{ij}\frac{(\pt_l\tilde{U})^2}{\tilde{U}^2}\right], \earr\eeq
where the derivatives are taken with respect to the cartesian
coordinates $\,\vec{x}~.$ The scalar curvature $\,R=4(Tr A)\,$
vanishes as it should.

The first equation in (\ref{EM}) gives for the field strength
$$F\equiv F^-+F^+=\pt_i\left[U^{-1}\right]\cdot\Xi^-_i
-\pt_i\left[{\tilde{U}^{-1}}\right]\cdot\Xi^+_i~.$$
Using (\ref{ipw2}),(\ref{ipw3}) one can check that these field
strengths indeed obey the Maxwell equations:
$$dF^+=dF^-=0~.$$

Let us prove the following :

\begin{nth}
The Perj\`es-Israel-Wilson metrics are self-dual Weyl only
in the two cases:
\brm
\item When $\,\tilde{U}\,$ is a constant : they are homothetic
to the multicentre metrics.
\item When $\,\tilde{U}=m/|\vec{x}-\vec{x}_0|\ $: they are
conformal to the multicentre metrics.
\erm
\end{nth}

\noindent{\bf Proof :} Let us impose, for instance, the condition that
the Weyl tensor is self-dual (i.e., $\,W^-=0$). Using (\ref{weyl1})
and (\ref{ipw2}), the corresponding constraints can be written as
\beq\label{fstep}\pt_i\partial_j\left(\frac 1{\tilde{U}^2}\right)-
\frac 13\de_{ij}\,\Delta\left(\frac 1{\tilde{U}^2}\right)=0~.\eeq
Acting on the left hand side by $\,\pt_i\,$ gives
$$\pt_j\,\Delta\left(\frac 1{\tilde{U}^2}\right)=0\qq\Longrightarrow\qq
\Delta\left(\frac 1{\tilde{U}^2}\right)=\mbox{const} \equiv 6B~.$$
Then one can integrate relation (\ref{fstep}) to
$$\left(\frac 1{\tilde{U}^2}\right)=A+\vec{f}\cdot\vec{x}+Br^2~,\qq
r^2= \vec{x}\cdot\vec{x}~,$$  where $\,A\,$ and $\,\vec{f}\,$ are integration constants.
The requirement that $\,\tilde{U}\,$ is harmonic (eq. \p{ipw2}) amounts to the
relation $$\,\vec{f}\cdot\vec{f}=4AB~.$$

If $\,B\,$ vanishes, the harmonic function $\,\tilde{U}\,$ is evidently reduced
to a constant  which can be scaled to 1. Then the relations (\ref{ipw2}),
(\ref{ipw3}) imply  that the metric  is homothetic to some multicentre one.

If $\,B\,$ does not vanish, we can write
$\dst\,\tilde{U}=m/|\vec{x}-\vec{x}_0|\,$ which can be simplified to $\,1/r$
by rescaling and translation of $\vec{x}~.$ The metric   (\ref{ipw1})
becomes  $$g=r^2\,\hat{g}~,$$  with  $$\hat{g}=\left[\frac
1{\hat{V}}(d\tau+{\cal A})^2 + \hat{V}\,\hat{\ga}\right],\qq
\hat{\ga}=\frac 1{r^4}\,{\hat \ga}_0~,\qq \hat{V}=r\,U~.$$
Using spherical coordinates we have
$${\hat
\ga}_0=dr^2+r^2\,d\Omega^2\qq\Longrightarrow\qq\hat{\ga}=d\rho^2+\rho^2\,d\Omega^2~,\qq
\rho=1/r~,$$
thus establishing that $\,\hat{\ga}\,$ is flat. Then relation (\ref{ipw3})
becomes  $$-\stdec_{\hat{\ga}}d{\cal A}=d\hat{V}~,$$
showing that $\,\hat{g}\,$ is some multicentre. This completes the proof.

Proposition 3 tells us that the metrics of Perj\`es-Israel-Wilson,
when they have self-dual Weyl tensor, are never conformal to Einstein
metrics (with non-vanishing cosmological constant). This implies that our
metric (\ref{met1}) can never be transformed to the Perj\`es-Israel-Wilson
form.
\subsection{The Plebanski-Demianski metric}
In \cite{pd} Plebanski and Demianski have derived a minkowskian solution of
the coupled Einstein-Maxwell field equations. Its euclidean version, obtained by complex changes of coordinates and parameters, can be written in the form  $\,\dst g_{PD}=\sum_{A=0}^3\ e_A^2,$ with the vierbein
$$\left\{\barr{ll}
\dst e_0=\frac 1{1+pq}\,\sqrt{\frac{p^2-q^2}{X(p)}}\,dp~, &
\dst e_1=\frac 1{1+pq}\,\sqrt{\frac{p^2-q^2}{Y(q)}}\,dq~,\\[4mm]
\dst e_2=\frac 1{1+pq}\,\sqrt{\frac{X(p)}{p^2-q^2}}\,(d\tau+q^2\,d\si)~, &
\dst e_3=\frac 1{1+pq}\,\sqrt{\frac{Y(q)}{p^2-q^2}}\,(d\tau+p^2\,d\si)~.
\earr\right.$$
where
$$\left\{\barr{l}
\dst X(p)=\left(g_0^2-\ga+\frac{\la}{6}\right)-2l\,p+\eps\,p^2-2m\,p^3-
\left(e^2+\ga+\frac{\la}{6}\right)\,p^4~,\\[4mm]
\dst Y(q)=\left(e^2+\ga-\frac{\la}{6}\right)-2m\,q-\eps\,q^2-2l\,q^3-
\left(g_0^2-\ga-\frac{\la}{6}\right)\,q^4~.\earr\right.$$
It displays 6 real parameters besides the cosmological constant $\,\la$
and possesses $U(1)\times U(1)$ isometry realized by shifts of $\tau$ and
$\sigma~.$

The meaning of the parameters $\,e,\,g_0,\,l\,$ and $\,m\,$ follows from:

\begin{nth}
The Plebanski-Demianski metrics are
\brm
\item[$\bullet$] Einstein for $\,e=g_0=0~.$
\item[$\bullet$] Einstein with self-dual Weyl tensor ($\,W^-=0\,$) for
$\,e=g_0=0\,$ and $\,l=m~.$
\item[$\bullet$] Einstein with anti-self-dual Weyl tensor ($\,W^+=0\,$) for
$\,e=g_0=0\,$ and $\,l=-m~.$
\erm
\end{nth}

\noindent{\bf Proof :} The proposition follows from the computation of the
curvature   matrices $\,A,B\,$ and $\,C$ defined in Appendix.

We are going to show that our metric \p{met1} lies outside the above ansatz. To
this end, we shall work with an anti-self-dual Weyl tensor ($\,W^-=0\,$)
and analyze the $\,\la\to 0\,$ limit of $g_{PD}~.$  We switch to  the
tri-holomorphic Killing vector $\,\partial_{\phi}\,$ by making the change of
coordinates   $$d\phi=d\tau~,\qq d\alf=d\si+d\tau~.$$  It leads to the
limiting metric   \beq\label{PD}
g_{PD}(\la\to 0)= \frac 1V(d\phi+{\cal A})^2+V\,\ga_0~,\eeq
with the potential
\beq\label{multi}
V=\frac{(1+pq)^2(p^2-q^2)}{\cal D}~,\qq {\cal
D}=(1-q^2)^2X(p)+(1-p^2)^2Y(q)~,\eeq  the gauge field one-form
\beq\label{gauge0}
{\cal A}=\frac{q^2(1-q^2)X(p)+p^2(1-p^2)Y(q)}{\cal D}\,d\alf~,\eeq
and the three dimensional metric
\beq\label{3dim}
\ga_0=\frac{\cal D}{(1+pq)^4}\left(\frac{dp^2}{X(p)}+
\frac{dq^2}{Y(q)}\right)+
\frac{X(p)Y(q)}{(1+pq)^4}\,d\alf^2~.\eeq
One can explicitly check the relation
$$\stdec_{\ga_0}\,d{\cal A}=\pm\,dV~.$$
To prove that (\ref{PD}) is indeed multicentre, we define cartesian
coordinates $\,\vec{x}\,$ by
\beq\label{pd3}
\left\{\barr{l}
x=A\,\sin\left[\sqrt{m^2+\ga(2\ga+\eps)}\,\alf\right],\\[4mm]
y=A\,\cos\left[\sqrt{m^2+\ga(2\ga+\eps)}\,\alf\right],\\[4mm]
z=B~,\earr\right.\eeq
with
$$A=\frac 1{(1+pq)^2}\sqrt{\frac{X(p)Y(q)}{m^2+\ga(2\ga+\eps)}}~,\quad
B=-\frac{m(p-q)(1-pq)+\ga(p^2+q^2)+\eps pq}
{\sqrt{m^2+\ga(2\ga+\eps)}(1+pq)^2}~.$$
One can check that these coordinates make manifest the flatness of the metric
(\ref{3dim})
$$\,\ga_0=(dx)^2+(dy)^2+(dz)^2~.$$

For comparing \p{PD} with the HK limit of our metric we need to
express the potential $\,V\,$ in terms of  the coordinates (\ref{pd3}).

For $\,m=0,$ as observed in the original paper \cite{pd}, the metric (\ref{PD})
is flat: this is   a special case which needs a separate analysis. We have
$$V=\frac 1{2\sqrt{\ga(\eps-2\ga)}}\frac 1{\sqrt{ x^2+y^2+Z^2}}~,\qq
Z=z+\frac{\eps}{2\sqrt{\ga(2\ga+\eps)}}~,$$
provided that the expressions within square roots are positive. This
potential corresponds to a mass at the origin, and is known to
yield a flat four-dimensional metric \cite{egh}.

For $\,m\neq 0\,$, we define new parameters by
$$\cosh\phi=\frac{\eps-2\ga}{4m}~,\qq\phi\geq 0~,\qq
c=\frac{\sqrt{(\eps-2\ga)^2-16m^2}}{4\sqrt{m^2+\ga(2\ga+\eps)}}~,$$
and
$$Z=z+\frac{2\ga+\eps}{4\sqrt{m^2+\ga(2\ga+\eps)}}~,
\qq d_{\pm}=x^2+y^2+(Z\pm c)^2~.$$
In this notation, the potential (\ref{multi}) becomes
\beq
\label{Vpd}
V=\mu\left(\frac{\eta}{\sqrt{d_-}}+
\frac{1/\eta}{\sqrt{d_+}}\right),
\eeq
with
$$ \eta^2=\frac{e^{-\phi}}{\sqrt{c^2+1}+c}~,\qq\qq
\mu=\frac{\sqrt{c}}{4m(\sinh\phi)^{3/2}}~.$$
The HK limit of the Plebanski-Demianski metric therefore gives
an ALE generalization of the Eguchi-Hanson metric (recovered for $\,\eta=1\,$)
with two different masses. It is reduced to the flat metric, up to rescaling,
in  the limits $\,\eta\to 0\,$ and $\,\eta\to\nf~.$

The potential (\ref{Vpd}) is a particular case $a = 0~,\; a\beta_0 \neq 0$ of
our limiting HK potential (\ref{pot1}):
$$V=\frac 14\left(a^2+\frac{1+a\be_0}{\sqrt{\de_-}}+
\frac{1-a\be_0}{\sqrt{\de_+}}\right).$$

The conclusion is that our general metric (\ref{met1}) cannot be embedded
into  the Plebanski-Demianski class of self-dual Einstein metrics because their
HK limits are different.\\

Summarizing the discussion in sections 6.3 and 6.4, we observe that our
metric \p{met1} cannot be reduced to either known class of metrics
solving the Einstein-Maxwell equations. Hence, by Flaherty's
equivalence, it provides (up to conformal factor \p{w}) a new family of explicit
solutions of this system. For the minimal case $\alpha_0 = \beta_0 = 0$ this
fact was pointed out in \cite{civ}.

\subsection{The linearization by Calderbank and Pedersen}
Quite recently, while we were typing this article,  Calderbank and Pedersen
\cite{cp} have exhibited a class of
self-dual Einstein metrics with two commuting (and hypersurface generating)
Killing vectors. To  describe their metrics, two main ingredients are needed:
\brm
\item A function  $\,F(\rho,\eta)\,$ which is a solution of the linear
differential equation
\beq\label{cp1}
\rho^2(F_{\rho\rho}+F_{\eta\eta})=\frac 34\,F~.\eeq
It is an eigenfunction of the laplacian in the hyperbolic plane
$\,{\cal H}^2\,$ with metric
\beq\label{cp2}
g_0({\cal H}^2)=\frac{d\rho^2+d\eta^2}{\rho^2}~,\qq\rho > 0~.\eeq
\item The set of one-forms
$$\alf=\sqrt{\rho}\,d\alf~,\qq\qq\be=\frac{d\phi+\eta\,d\alf}{\sqrt{\rho}}~.$$
\erm
In terms of these, the full metric is
\beq\label{cp3}
\barr{ll}
\dst g=\frac{F^2-4\rho^2(F_{\rho}^2+F_{\eta}^2)}{4F^2}\,g_0({\cal H}^2)\\[5mm]
\dst
\hspace{3cm}+\,\frac{[(F-2\rho\,F_{\rho})\,\alf-2\rho\,F_{\eta}\,\be)^2+(2\rho\,F_{\eta}
\,\alf-(F+2\rho\,F_{\rho})\,\be]^2}{F^2[F^2-4\rho^2(F_{\rho}^2+F_{\eta}^2)]}~.\earr\eeq
The main result of \cite{cp} is a theorem which states that
these metrics with two commuting Killings are self-dual Einstein with
non-vanishing scalar curvature  and that {\em any} such metric has locally the
structure given by the  expression (\ref{cp3}).

In order to get a deeper insight into the construction of \cite{cp},
it is convenient to pass to a  function $\,G\,$ according to
$\,F=G/\sqrt{\rho}~.$ The metric $g$ becomes \beq\label{caldped} G^2\,g=\frac
1{\cal W}(d\phi+\Theta)^2+{\cal W}\,\ga~,\qq\qq
\Theta=\left(\frac{G\,G_{\eta}}{G_{\rho}^2+G_{\eta}^2}-\eta\right)d\alf~,\eeq
with $${\cal W}=\frac 1\rho\,\frac{G\,G_{\rho}}{G_{\rho}^2+G_{\eta}^2}-1~,\qq
\ga=\rho^2\,d\alf^2+(G_{\rho}^2+G_{\eta}^2)\,\left(d\rho^2+d\eta^2\right).$$

Following Tod, we can now compute the anti-self-dual part of $\,dK,$ where
$\,K\,$ is
the 1-form associated with the Killing $\,\pt_{\phi}~.$ After some algebra,
using (\ref{cp1}), we obtain
$$K=\frac 1{G^2\,{\cal W}}(d\phi+\Theta)~,\qq\quad dK^-=-\frac
1{G\sqrt{G_{\rho}^2+G_{\eta}^2}}
\left(G_{\rho}\ \Xi^-_1+G_{\eta}\ \Xi^-_2\right),$$
{}from which we conclude that in fact $\,G\,$ is proportional to Tod's
coordinate $\,w,$ defined in (\ref{sc}).
Taking $\,G=w,$ relation (\ref{cp1}) becomes
\beq\label{cp4}
w_{\rho\rho}+w_{\eta\eta}=\frac 1{\rho}\,w_{\rho}~.\eeq

Using relations (\ref{omega}) and (\ref{rho}), and switching
from Tod's coordinates $(w,\nu)$ to the $(\rho,\eta)$ coordinates, we can
obtain the K\"ahler form $\Omega$ and the Ricci form ${\hat \rho}$ in this
setting:
\beqa
\Omega &=& -dw\wedge
d\phi+\left(\eta\,w_\rho-\rho\,w_\eta\right)\,d\rho\wedge
d\alpha+\left(\rho\,w_\rho+\eta\,w_\eta-w\right)\,d\eta\wedge d\alpha\ ,\nn
{\hat \rho} &=&\dst -d\left[\frac1{w\,{\cal
W}}\left(d\phi+\Theta\right)+\frac1w\left(d\phi-\eta\,d\alpha\right)\right]\ .
\label{omerho}
\eeqa
The K\"ahler form $\Omega$ is closed as a consequence of (\ref{cp4}). One
can check that $\Omega$ and $\hat \rho$ possess opposite self-dualities.
In view of Flaherty's equivalence, the metrics described
by the Calderbank-Pedersen ansatz are conformally related to a subclass of
metrics solving the coupled Einstein-Maxwell equations. Then the
two-forms \p{omerho} specify the field strengths of the corresponding Maxwell
field (\ref{max}).

We are now in a position to establish the precise connection between our
coordinates $s$ and $x$ and the coordinates $\rho$  and $\eta$ in the
hyperbolic plane ${\cal H}^2$. To this end, we have to identify  the pieces
which are independent of the choice of basis for the Killing  vectors, i.e. the
pieces involving $\gamma$. One gets the correspondence:
\beq
\begin{array}{rll}
\label{rhoeta}
\dst \rho &=&\dst \frac{4\,s}{\delta({\hat c})}\,{\sqrt{1 - x^2}}\,{\sqrt{1 +
a^2\,s^2\,\lambda }}~,\\[5mm]
\dst \eta &=&\dst \frac{2}{{\hat c}} \left(\frac{s\,\left( s +2\,{\hat c}\,x
\right) }{\delta({\hat c})}-1\right),
\end{array}
\eeq
where $\delta({\hat c})$ was defined in (\ref{delta}) and ${\hat
c}=\alpha_0\,c/a~.$ Let us notice that the coordinate $\eta$ is defined up
to an additive constant that can always be re-absorbed through a redefinition
of the Killing $\pt_\phi$. The check of equation (\ref{tod}a) gives
$\,\Lambda=3 \Leftrightarrow\lambda=-1/16$. One can then invert relations
(\ref{rhoeta}) :
$$
\begin{array}{rll}
\dst x &=&\dst \frac{\left| 2 - c\,\alpha_0  \right| \,\sqrt{\left(\eta
-\frac{2\,a}{2 - c\,\alpha_0 } \right)^2 + \rho^2} -
    \left| 2 + c\,\alpha_0  \right| \,\sqrt{\left(\eta +\frac{2\,a}{2 +
c\,\alpha_0 } \right) ^2 + {\rho }^2}}
{2\,\sqrt{\left(c\,\alpha_0\,\eta+2\,a \right)^2 + c^2\,\alpha_0^2\,{\rho}^2}}
\,\\[10mm]
\dst s &=&\dst 8\,\frac{{\sqrt{\left(\frac{c\alpha_0}{a}\,\eta+
2\right)^2 + \left(\frac{c\alpha_0}{a}\right)^2\,{\rho }^2}}}
  {\left| 2 - c\,\alpha_0  \right| \,\sqrt{\left(\eta -\frac{2\,a}{2 -
c\,\alpha_0 } \right)^2 + \rho^2} +
    \left| 2 + c\,\alpha_0  \right| \,\sqrt{\left(\eta +\frac{2\,a}{2 +
c\,\alpha_0 } \right) ^2 + {\rho }^2}}\ .\\[10mm]
\end{array}
$$

As discussed in section 5.5, in the analysis of the QK Eguchi-Hanson
limit, for $\,a\to 0\,$ the parameter $\,\alpha_0\,$ becomes irrelevant since
it disappears from the metric. The above coordinate $s$ is well defined
in the limit $\,a\to 0\,$ only if we first put $\alpha_0 = 0.$

Having the explicit expressions for $s, x$, it is then possible to compute
$w(\rho,\eta)$ which was given in (\ref{w}) as a function of $s,x $:
$$
\begin{array}{rll}
\dst w &=&\dst \frac{1}{4}\,\left| 2 - c\,\alpha_0  \right|
\,{\sqrt{{\left(\eta-\frac{2\,a}{2 - c\,\alpha_0 }\right) }^2 + {\rho }^2}} +
  \frac{1}{4}\,\left| 2 + c\,\alpha_0  \right|
\,{\sqrt{{\left(\eta+\frac{2\,a}{2 + c\,\alpha_0 } \right) }^2 +{\rho
}^2}}\\[5mm]
&&\dst +
  \frac{|c|}{8}\,{\sqrt{{\left(\eta-\frac{2}{c}\,\left( 1 - a\,\beta_0  \right)
\right) }^2 + {\rho }^2}} +
  \frac{|c|}{8}\,{\sqrt{{\left(\eta+\frac{2}{c}\,\left( 1 + a\,\beta_0  \right)
\right) }^2 + {\rho }^2}}~.
\end{array}
$$
It is easy to check that $w(\rho,\eta)$ satisfies eq. (\ref{cp4}).

Let us finally give $w(\rho,\eta)$ in the two interesting cases
$a\rightarrow0$ (QK-EH) and $c\rightarrow0$ (QK-TN):
$$
\begin{array}{rll}
\dst w_{\hbox {\tiny QK-EH}}(\rho,\eta) &=& \dst \sqrt{\eta^2+\rho^2} +
\frac{|c|}{8}\,\sqrt{\left(\eta-\frac{2}{c}\right)^2 + \rho^2} +
\frac{|c|}{8}\,\sqrt{\left(\eta+\frac{2}{c}\right)^2 + \rho^2}~,\\[6mm]
\dst w_{\hbox {\tiny QK-TN}}(\rho,\eta) &=& \dst \frac{1}{2} +
\frac{1}{2}\,\sqrt{\left(\eta-a\right)^2 + \rho^2} +
  \frac{1}{2}\,\sqrt{\left(\eta+a\right)^2 +\rho^2}~.
\end{array}
$$

Using these relations we can, e.g., compute the forms $\Omega $ and $\hat\rho $
\p{omerho} for our metrics and, via the correspondence \p{max}, to find the
relevant Maxwell field strengths.

\setcounter{equation}{0}
\section{Conclusions}
In this paper, proceeding from the general HSS formulation of QK sigma models,
we have constructed a wide class of $U(1)\times U(1)$ 4-dimensional
QK metrics extending most general two-centre HK metrics. These QK metrics
supply, via Flaherty's equivalence \cite{Fl}, a new family of explicit
solutions of the coupled Einstein-Maxwell equations. We have given the precise
embedding  of our metrics in the framework of general $U(1)\times U(1)$ ansatz
of Calderbank and Pedersen \cite{cp}.

The HSS approach gives QK metrics in the form which admits a transparent
interpretation of the involved parameters as the symmetry breaking ones and
possesses a clear hyper-K\"ahler limit, with the Einstein
constant as a contraction parameter. However, despite these attractive
features, it does not immediately provide the natural coordinates best suited
to display the final linearization of the self-dual Einstein equations along
the line of ref. \cite{cp}. It would be interesting to explore what the
choice of such coordinates means in the language of the original
hypermultiplet superfields  parametrizing the general HSS action of QK sigma
models. One more obvious direction  of further study  could consist in
applying our HSS methods for explicit construction of higher-dimensional QK
metrics generalizing, e.g., the HK metrics constructed  in \cite{ggr}.

One of possible physical applications of the QK metrics presented here is to
utilize them in the context of gauged five-dimensional supergravities. The
latter seemingly provide an appropriate framework for supersymmetric
extensions of the famous Randall-Sundrum scenario (for a recent review, see
\cite{Fre}). The presence of matter hypermultiplets seems necessary for the
existence of such (smooth) extensions (see, e.g., \cite{cdkv}). To analyse
various possibilities, it is important to know the structure of the scalar
potential which is obtained by gauging isometries of the QK manifold
parametrized  by the hypermultiplets. Until now, in the actual computations
(e.g., in  \cite{flp,cdkv}) there was mainly used the so-called universal
hypermultiplet \cite{1} corresponding to the homogeneous QK manifold  $SU(2,
1)/U(2)$. It would be tempting  to study models with non-homogeneous QK
manifolds possessing isometries and, in particular, with those considered here.
It is straightforward to gauge the $U(1)\times U(1)$ isometries of our HSS
actions following the general recipe of ref. \cite{bgio} (in order to
generate scalar potentials, the relevant gauge supermultiplets should be
propagating, in contrast to the non-propagating ones employed in the HSS
quotient). The  $SU(2,1)/U(2)$ QK manifold is a special case \cite{IvV} of the
QK Eguchi-Hanson limit of our $U(1)\times U(1)$ class of QK manifolds, so  the
scalar potentials associated with our metrics and inheriting all free
parameters of the latter may offer new possibilities as compared to the case
of universal hypermultiplet.

\vspace{1.3cm}

\noindent{\Large\bf Acknowledgements}\\

\noindent E.I. thanks Directorate of Laboratoire de Physique Th\'eorique et
des Hautes Energies, Universit\'e Paris VII, for the hospitality
extended to him during the course of this work under the Project PAST-RI 99/01.
His work was partially supported by the grants RFBR 99-02-18417,  RFBR-CNRS
98-02-22034, INTAS-00-0254, NATO Grant PST.CLG  974874 and PICS Project No.
593. \vspace{1cm}

\def\theequation{A.\arabic{equation}}
\setcounter{equation}{0}
\section*{Appendix. Definitions and notation}

For a given metric $\,g,$ the vierbein $\,e_a,\ a=0,1,2,3,$ is such that
$\dst\,g=\sum_a\,e_a^2~.$ The spin connection $\,\om_{ab}\,$ is defined by 
$$de_a+\om_{ab}\wedge e_b=0~,\qq\qq \om_{ab}=-\om_{ba}~,$$ 
with self-dual components 
$$\om^{\pm}_i=\om_{0i}\pm\frac 12\,\epsilon_{ijk}\,\om_{jk}~,$$ 
and similarly for the curvature 
$$R_{ab}=d\om_{ab}+\om_{as}\wedge\om_{sb}=
\frac 12\,R_{ab,st}\,e_s\wedge e_t~,\quad\rightarrow\quad  
R^{\pm}_i=R_{0i}\pm\frac 12\,\epsilon_{ijk}\,R_{jk}~.$$ 
We take for the Ricci tensor and scalar curvature 
$$Ric_{\,ab}=R_{as,bs}~,\qq\qq R=Ric_{\,ss}.$$ 
It is useful to define the two-forms of definite self-duality by 
$$\Xi^{\pm}_i=e_0\wedge e_i\pm \frac 12\,\epsilon_{ijk}\,e_j\wedge e_k~.$$ 
Using this basis, the curvature and Ricci tensor are encoded in the three  
matrices $\,A,\,B\,$ and $\,C\,$ such that 
$$R^+_i=A_{ij}\ \Xi^+_j+B_{ij}\ \Xi^-_j~,\qq R^-_i=B^t_{ij}\ \Xi^+_j+C_{ij}\
\Xi^-_j~,$$ 
where the matrices $\,A\,$ and $\,C\,$ are symmetric.  
 
The Ricci components in the vierbein basis are  
$$Ric\,_{00}=Tr(A+B)~,\quad Ric\,_{0i}=-\frac
12\epsilon_{ijk}\,(B_{jk}-B^t_{jk})~,\quad  
Ric\,_{ij}=Tr(A-B)\,\de_{ij}+B_{ij}+B^t_{ij}~,$$ 
and the scalar curvature is
$$ R=4\,(Tr A)=4\,(Tr C)~.$$ 
The Einstein condition $\, Ric_{\,ab}=\Lambda\,\de_{ab}\,$ is seen to be 
equivalent to the vanishing of the matrix $\,B\,$ and we have $\,Tr\, C= Tr\,
A=\Lambda~.$
 
One further defines the Weyl tensor 
$$W_{ab,cd}=R_{ab,cd}+\frac R6(\de_{ac}\,\de_{bd}-\de_{ad}\,\de_{bc})- 
\frac
12(\de_{ac}\,Ric_{bd}-\de_{ad}\,Ric_{bc}+\de_{bd}\,Ric_{ac}-\de_{bc}\,Ric_{ad})~.$$ 
The corresponding two-forms  
$$W_{ab}=\frac 12\,W_{ab,cd}\,e_c\wedge e_d~,$$ 
and their self-dual parts are given by 
$$\left\{\barr{ll}W^+_i\equiv W_{0i}+\frac 12\,\eps_{ijk}\,W_{jk}=W^+_{ij}\
\Xi^+_j~,  
& \qq W^+_{ij}=A_{ij}-\frac 13\,(Tr A)\,\de_{ij}~,\\[4mm] 
W^-_i\equiv W_{0i}-\frac 12\,\eps_{ijk}\,W_{jk}=W^-_{ij}\ \Xi^-_j~, 
& \qq W^-_{ij}=C_{ij}-\frac 13\,(Tr C)\,\de_{ij}~.\earr\right.$$ 
We conclude that for an Einstein space with self-dual Weyl tensor (i. e.,
$\,W^-_i=0\,$) 
we should have
$$ C_{ij}=\frac{\Lambda}{3}\ \de_{ij}\qq\Longleftrightarrow\qq
R^-_i=\frac{\Lambda}{3}\ \Xi^-_i~.$$

\end{document}